\documentclass[12pt]{iopart}
\usepackage{enumerate}
\usepackage{graphicx} 
\usepackage{multirow}
\usepackage{array} 
\newtheorem{thmn}{Theorem}
\usepackage{hyperref} 

\begin{document}


\title[Symmetries]{Symmetries of nonlinear ordinary differential equations: the modified Emden equation as a case study}

\author{M. Senthilvelan$^1$, V. K. Chandrasekar$^2$, R. Mohanasubha$^1$}
\address{$^1$Centre for Nonlinear Dynamics, School of Physics,
Bharathidasan University, Tiruchirappalli - 620 024, India}
\address{$^2$Centre for Nonlinear Science and Engineering, School of Electrical and Electronics Engineering, SASTRA University, Thanjavur - 613 401, India}
 
\begin{abstract}
Lie symmetry analysis is one of the powerful tools to analyze nonlinear ordinary differential equations. We review the effectiveness of this method in terms of various symmetries. We present the method of deriving Lie point symmetries, contact symmetries, hidden symmetries, nonlocal symmetries, $\lambda$-symmetries, adjoint symmetries and telescopic vector fields of a second-order ordinary differential equation. We also illustrate the algorithm involved in each method by considering a nonlinear oscillator equation as an example. The connections between (i) symmetries and integrating factors and (ii) symmetries and integrals are also discussed and illustrated through the same example. The interconnections between some of the above symmetries, that is (i) Lie point symmetries and $\lambda$-symmetries and (ii) exponential nonlocal symmetries and $\lambda$-symmetries are also discussed. The order reduction procedure is invoked to derive the general solution of the second-order equation.
\end{abstract}

 
\maketitle

\section{Introduction}
During the past four decades a great deal of interest has been shown in identifying and characterizing qualitative and quantitative properties of finite dimensional integrable nonlinear dynamical systems \cite{book1,int12}. Several powerful mathematical methods have been developed to solve/integrate nonlinear ordinary differential equations (ODEs). Some widely used methods to solve nonlinear ODEs in the contemporary literature are (i) Painlev$\acute{e}$ singularity structure analysis \cite{pain,pain1,mlbook}, (ii) Lie symmetry analysis \cite{olv,hydon,stephani,blu1,blu,ibr}, (iii) Darboux method \cite{darb1} and (iv) Jacobi last multiplier method \cite{jlm1,nuci1,nuci}. Among these the Lie group method advocated by Sophus Lie plays a vital role \cite{olv,hydon,stephani,blu1,ibr}. The method is essentially based on the invariance of differential equations under a continuous group of point transformations and such transformations usually have the form $T=T(t,x,\epsilon),~X=X(t,x,\epsilon),$ where $(t,x)$ and $(T,X)$ are the old and new independent and dependent variables respectively of the given ODE and $\epsilon$ denotes the group parameter. The transformations depend only on the variables $t$ and $x$ and not on the derivatives, that is $\dot{x}$. Transformations of this type are called the point symmetry group of a differential equation when this group of transformations leave the differential equation invariant \cite{hydon,stephani,blu1,blu}. The Norwegian mathematician, Sophus Lie, founder of this method, developed an algorithm to determine the symmetry groups associated with a given differential equation in a systematic way. Once the symmetry group associated with the differential equation is explored, it can be used to analyze the differential equation in several ways. For example, the symmetry groups can be used (i) to derive new solutions from old ones \cite{blu1,olv}, (ii)  to reduce the order of the given equation \cite{olv,blu1,hydon}, (iii) to discover whether or not a differential equation can be linearized and to construct an explicit linearization when one exists \cite{kum1,kum2,new_leach} and (iv) to derive conserved quantities \cite{olv}. 

However, studies have also shown that certain nonlinear ODEs which are integrable by quadratures do not admit Lie point symmetries \cite{lopex,mur1}. To understand the integrability of these nonlinear ODEs, through Lie symmetry analysis, attempts have been made to extend Lie's theory of continuous group of point transformations in several directions. A few notable extensions which have been developed for this purpose are (i) contact symmetries \cite{cont,ci4,ci5,ci3}, (ii) hidden and nonlocal symmetries \cite{ci6,non_past,nonlocal,nonlocal1,nonlocal2,nonlocal3,nonlocal4,nonlocal5,nonlocal7,si11,si12,sir2}, (iii) $\lambda$-symmetries \cite{mur1,mur3,mur4,tel2}, (iv) adjoint symmetries \cite{blu,blu1,blu11} and (v) telescopic vector fields \cite{tel1,tel2}.

In the conventional Lie symmetry analysis the invariance of differential equations under one parameter Lie group of continuous transformations is investigated with point transformations alone. One may also consider the coefficient functions $\xi$ and $\eta$ (see Eq.(2.2)) in the infinitesimal transformations to be functions of $\dot{x}$ besides $t$ and $x$. Such derivative included transformations are called contact transformations. In fact, Lie himself considered this extension \cite{19}. The method of finding contact symmetries for certain linear oscillators (harmonic and damped harmonic oscillators) were worked out by Schwarz and Cerver$\acute{o}$ and Villarroel \cite{ci4,ci7}. The integrability of a class of nonlinear oscillators through dynamical symmetries approach was carried out by Lakshmanan and his collaborators, see for example Refs. \cite{sir_old1,sir_old2}.

Investigations have also revealed that nonlinear ODEs do admit nonlocal symmetries. A symmetry is nonlocal if the infinitesimal transformations depend upon an integral. Initially some of these nonlocal symmetries were observed as hidden symmetries of ODEs in the following manner. Suppose an $n^{th}$-order ODE is order reduced to $(n-1)^{th}$ order with the help of a Lie point symmetry. Now substituting the transformation (which was used to reduce the $n^{th}$-order to $(n-1)^{th}$ order) in other symmetry vector fields of the $n^{th}$-order equation one observes that these symmetry vector fields turns out to be symmetry vector fields of the order reduced ODE. In other words, all these vector fields satisfy the linearized equation of the order reduced equation. Upon analyzing these vector fields one may observe that some of them retain their point symmetry nature and the rest of them turn out to be nonlocal vector fields of the reduced ODE. Since these nonlocal symmetry vector fields cannot be identified through Lie point symmetry analysis they are coined as hidden symmetries. These nonlocal hidden symmetries were first observed by Olver and later they were largely investigated by Abraham-Shrauner and her collaborators \cite{nonlocal1, nonlocal3,nonlocal5}. 

Subsequently it has been shown that many of these nonlocal or hidden symmetries can be connected to $\lambda$-symmetries. The $\lambda$-symmetries concept was introduced by Muriel and Romero \cite{mur1}. These $\lambda$-symmetries can be derived by a well-defined algorithm which include Lie point symmetries as a specific sub-case and have an associated order reduction procedure which is similar to the classical Lie method of reduction \cite{mur3}. Although, $\lambda$-symmetries are not Lie point symmetries, the unique prolongation of vector fields to the space of variables $(t, x, \dot{x},\ddot{x},...)$ for which the Lie reduction method applies is always a $\lambda$-prolongation, for some functions $\lambda(t, x, \dot{x},\ddot{x},...)$. For more details on $\lambda$-symmetries approach, one may refer the works of Muriel and Romero \cite{mur4}. The method of finding $\lambda$-symmetries for a second-order ODE has been discussed in depth by these authors and the advantage of finding such symmetries has also been demonstrated by them. The authors have also developed an algorithm to determine integrating factors and integrals from $\lambda$-symmetries for second-order ODEs \cite{mur3}.

Very recently Pucci and Sacomandi have generalized $\lambda$-symmetries by introducing telescopic vector fields. Telescopic vector fields are more general vector fields which encompose Lie point symmetries, contact symmetries and $\lambda$-symmetries as their sub-cases. For more details about these generalized vector fields we refer to the works of Pucci and Sacomandi \cite{tel1}.

The connection between symmetries and integrating factors of higher order ODEs was investigated by several authors \cite{ci8}. The literature is large and in this paper we discuss only one method, namely adjoint symmetry method which was developed by Bluman and Anco \cite{blu,blu1}. The main observation of Bluman and Anco was that the integrating factors are the solutions of adjoint equation of the linearized equation. In case the adjoint equation coincides with the linearized equation, then the underlying system is self-adjoint and in this case the symmetries become the integrating factors. A main advantage of this method is that we can find the integrals straightaway by multiplying the integrating factors and integrating the resultant ODE \cite{blu11}.

The symmetry methods described above are all applicable to any order. Each method has its own merits and demerits. In this paper, we review the methods of finding symmetries (starting from Lie point symmetries to telescopic vector fields) of a differential equation and demonstrate how these symmetries are helpful in determining integrating factors and integrals of the ODEs and establish the integrability of them. We demonstrate all these symmetry methods for a second-order ODE. The extension of each one of this procedure to higher order ODEs is effectively an extension. We also illustrate each one of the methods with the same example. We consider the same example for all the methods so that the general reader can understand the advantages, disadvantages and complexity involved in each one of these methods. 

The example which we have chosen to illustrate the symmetry methods is the modified Emden equation (MEE) \cite{exboo2,int14,int15,Dix:1990,int17,chand1,carinena,ames}
\begin{equation}
\ddot{x}+3{x}\dot{x}+x^3=0,\qquad \qquad.=\frac{d} {dt}. \label{main} 
\end{equation}
In the contemporary literature Eq. (\ref{main}) are also called second-order Riccati equation /Painlev$\acute{e}$-Ince equation. This equation has received attention from both mathematicians and physicists for a long time \cite{int1,int12}. For example, Painlev$\acute{e}$ had studied this equation with two arbitrary parameters, $\ddot{x}+\alpha{x}\dot{x}+\beta x^3=0$, and identified Eq. (\ref{main}) as one of the four integrable cases of it \cite{pain}. The differential equation (\ref{main}) arises in a variety of mathematical problems such as univalued functions defined by second-order differential equations \cite{exboo1} and the Riccati equation \cite{exboo2}. On the other hand, physicists have shown that this equation arises in different contexts. For example, it arises in the study of equilibrium configurations of a spherical gas cloud acting under the mutual attraction of its molecules and subjected to the laws of thermodynamics \cite{int15,Dix:1990}. Equation (\ref{main}) admits time independent nonstandard Lagrangian and Hamiltonian structures \cite{chand1}. In the contemporary literature, this equation has been considered by several authors in different contexts. For example, Chandrasekar, Senthilvelan and Lakshmanan have studied the linearization and investigated the integrability of this equation through the extended Prelle-Singer procedure \cite{chand1}. The Lie point symmetries of this equation were also derived by few authors in different contexts. For example, Mahomed and Leach have studied the invariance of this equation and shown that it admits $sl(3,R)$ algebra and constructed a linearizing transformation from the Lie point symmetries. Pandey et al. have identified (\ref{main}) as one of the nonlinear ODEs that admits maximal Lie point symmetries when they carry out Lie symmetry analysis for the equation, $\ddot{x}+f(x)\dot{x}+g(x)=0$, where $f(x)$ and $g(x)$ are functions of $x$ \cite{pandey1}. Nucci and her group have analyzed this equation in terms of Jacobi last multiplier \cite{nuci}. $\lambda$-symmetries and their associated integrating factors for this equation were investigated by Bhuvaneswari et.al. \cite{bhu1}. Noether symmetries of this equation were also studied in Ref. \cite{arxiv}. The nonlocal symmetries were also investigated in the Refs. \cite{nonlocal3,nonlocal} 

The plan of the paper is as follows. In Sec. \ref{2}, we present Lie's invariance analysis to determine Lie point symmetries of Eq.(\ref{main}). We also discuss few applications of Lie point symmetries. In Sec. \ref{noeth}, we describe the method of finding variational symmetries of this equation. In Sec. \ref{comta}, we consider a more generalized transformation and present the method of finding contact symmetries of the given differential equation. In Sec. \ref{sec5}, we discuss the methods that connect symmetries and integrating factors. We consider two different methods, namely, (i) adjoint symmetries method and (ii) $\lambda$-symmetries approach. In Sec. \ref{hidd}, we introduce the notion of hidden symmetries and enlist some hidden symmetries of MEE which are not obtained through Lie symmetry analysis. In Sec. \ref{nongan}, we introduce nonlocal symmetries and investigate  the connection between nonlocal symmetries and $\lambda$-symmetries. In Sec. \ref{teles}, we consider a more general vector field, namely telescopic vector field and derive these generalized vector fields for the MEE. Finally, we give our conclusions in Sec. \ref{9th}.

\section{Lie point symmetries}
\label{2}

Let us consider a second-order ODE 
\begin{eqnarray}
\ddot{x}=\phi(t,x,\dot{x}),\label{main1}
\end{eqnarray}
where overdot denotes differentiation with respect to `$t$'. The invariance of Eq.(\ref{main1}) under an one parameter Lie group of infinitesimal point transformations \cite{mahomed1,pandey1},
\begin{eqnarray}
&&T=t+\varepsilon \,\xi(t,x),~~~X=x+\varepsilon \,\eta(t,x),\quad \epsilon \ll 1,\label{asjkm}
\end{eqnarray}
where $\xi(t,x)$ and $\eta(t,x)$ are arbitrary functions of their arguments and $\varepsilon$ is a small parameter, is given by
\begin{eqnarray}
&&\hspace{-1.9cm}\xi \frac{\partial \phi}{\partial t}+\eta \frac{\partial \phi}{\partial x}+(\eta_t+\dot x (\eta_x-\xi_t)-\dot x^2 \xi_x)\frac{\partial \phi}{ \partial \dot x}-(\eta_{tt}+(2\eta_{tx}-\xi_{tt})\dot x+(\eta_{xx}-2\xi_{tx})\dot x^2\nonumber \\&& \hspace{4.30cm}-\xi_{xx} \dot x^3+(\eta_x-2\xi_t-3\dot x \xi_x)\ddot x) =0.\label{liec1}
\end{eqnarray}
Substituting the known expression $\phi$ in (\ref{liec1}) and equating the coefficients of various powers of $\dot{x}$, to zero one obtains a set of linear partial differential equations for the unknown functions $\xi$ and $\eta$. Solving them consistently we can get the Lie point symmetries ($\xi$ and $\eta$) associated with the given ODE. The associated vector field is given by $V=\xi \frac{\partial} {\partial t}+\eta\frac{\partial} {\partial x}$. 

One may also introduce a characteristics $Q=\eta-\dot{x}\xi$ and rewrite the invariance condition (\ref{liec1}) in terms of a single variable $Q$ as \cite{olv}
\begin{equation}
\frac{d^2Q} {dt^2}-\phi_{\dot{x}}\frac{dQ} {dt}-\phi_x Q=0.
\label{met1411}
\end{equation}
Solving Eq.(\ref{met1411}) one can get $Q$. From $Q$ one can recover the functions $\xi$ and $\eta$. The invariants associated with the vector field $V$ can be found by solving the following characteristic equation \cite{olv,blu1}
\begin{equation}
\frac{dt} {\xi}=\frac{dx} {\eta}=\frac{d\dot{x}}{\eta^{(1)}}.\label{invaria}
\end{equation}
Here $\eta^{(1)}$ represents the first prolongation which is given by $\eta_t+\dot x (\eta_x-\xi_t)-\dot x^2 \xi_x$. Integrating the characteristic equation (\ref{invaria}) we obtain two invariants, namely $u(t,x)$ and $v(t,x,\dot{x})$. The derivative of these two invariants, 
\begin{eqnarray}
w=\frac{dv} {du}=\frac{v_t+\dot{x}v_x+\phi v_{\dot{x}}} {u_t+\dot{x}u_x},\label{cite1}
\end{eqnarray} 
defines a second-order differential invariant. Integrating the above equation (\ref{cite1}) we can get the solution for the given equation (\ref{main1}).
\subsection{Example: modified Emden equation}
Eq.(\ref{main}) is invariant under the infinitesimal transformation (\ref{asjkm}) provided it should satisfy the Eq.(\ref{liec1}). Substituting the expression $\phi=-3 x \dot{x}-x^3$ in (\ref{liec1}), we get
\begin{eqnarray}
\hspace{-1cm}\eta(-3\dot{x}&-3x^2)+(\eta_t+\dot{x}\eta_x-\dot{x}\xi_t-\dot{x}^2\xi_x)(-3x)-(\eta_{tt}+\dot{x}(2\eta_{tx}-\xi_{tt})\nonumber \\ 
&\hspace{-0.9cm}+\dot{x}^2(\eta_{xx}-2\xi_{tx})-\xi_{xx}\dot{x}^3+(\eta_x-2\xi_t-3\dot{x}\xi_x)(-3 x \dot{x}-x^3))=0.
\end{eqnarray}
Equating the coefficients of various powers of $\dot{x}$ to zero and solving the resultant set of partial differential equations for $\xi$ and $\eta$, we obtain \cite{bhu1,pandey1,new_leach}
\begin{eqnarray}
\xi&=&x\bigg(a_2+a_1t-\frac{c_2+b_1} {2}t^2-\frac{c_1+d_2} {2}t^3+\frac{d_1} {4}t^4\bigg)\nonumber \\&&-\frac{d_1} {2}t^3+\bigg(c_1+\frac{3} {2} d_2\bigg)t^2+b_1t+b_2,\nonumber\\
\eta&=&-x^3\bigg(a_1t+a_2+\frac{d_1} {4}t^4-\bigg(\frac{c_1+d_2} {2}\bigg)t^3-\frac{c_2+b_1} {2}t^2\bigg)\nonumber \\&&+x^2\bigg(d_1t^3-3\bigg(\frac{c_1+d_2} {2}\bigg)t^2-(c_2+b_1)t+a_1\bigg)\nonumber \\&&+x\bigg(-\frac{3} {2} d_1t^2+c_1t+c_2\bigg)+d_1t+d_2,
\end{eqnarray}
where $a_i,b_i,c_i$ and $d_i,~i=1,2,$ are real arbitrary constants. The associated Lie vector fields are given by
\begin{eqnarray}
&&V_1=\frac{\partial} {\partial{t}},~~V_2=t\bigg(1-\frac{xt} {2}\bigg)\frac{\partial} {\partial{t}}+x^2t\bigg(-1+\frac{xt} {2}\bigg)\frac{\partial} {\partial{x}},\nonumber\\
&&V_3=x\frac{\partial} {\partial{t}}-x^3\frac{\partial} {\partial{x}},~~V_4=xt\frac{\partial} {\partial{t}}+x^2\bigg(1-xt\bigg)\frac{\partial} {\partial{x}},\nonumber\\
&&V_5=-\frac{xt^2} {2}\frac{\partial} {\partial{t}}+x\bigg(1-xt+\frac{x^2 t^2} {2}\bigg)\frac{\partial} {\partial{x}},\nonumber\\
&&V_6=t^2\bigg(1-\frac{xt} {2}\bigg)\frac{\partial} {\partial{t}}+xt\bigg(1-\frac{3} {2}xt+\frac{x^2 t^2} {2}\bigg)\frac{\partial} {\partial{x}},\nonumber\\
&&V_7=\frac{3} {2}t^2\bigg(1-\frac{xt} {3}\bigg)\frac{\partial} {\partial{t}}+\bigg(1-\frac{3} {2}x^2t^2+\frac{x^3 t^3} {2}\bigg)\frac{\partial} {\partial{x}},\nonumber\\
&&V_8=-\frac{t^3} {2}\bigg(1-\frac{xt} {2}\bigg)\frac{\partial} {\partial{t}}+t\bigg(1-\frac{3} {2}xt+x^2t^2-\frac{x^3t^3} {4}\bigg)\frac{\partial} {\partial{x}}.\label{vf8}
\end{eqnarray}

One can explore the algebra associated with the Lie group of infinitesimal transformations (\ref{asjkm}) by analyzing the commutation relations between the vector fields. Since the example under consideration admits maximal Lie point symmetries (eight) the underlying Lie algebra turns out to be $sl(3,R)$ which can be unambiguously verified from the vector fields (\ref{vf8}) \cite{new_leach}. In the following, we present few applications of Lie vector fields (\ref{vf8}).

\subsection{Applications of Lie point symmetries}
\subsubsection{General solution}
\label{sol_lie}
The first and foremost application of Lie point symmetries is to explore the solution of the given equation through order reduction procedure. The order reduction procedure is carried out by constructing the invariants associated with the vector fields \cite{olv,blu1}. In the following, we illustrate the order reduction procedure by choosing the vector field $V_3$. For the remaining vector fields one can proceed and obtain the solution in the same manner.

Substituting the expressions $\xi$, $\eta$ and $\eta^{(1)}$ in the characteristic equation $\frac{dt} {\xi}=\frac{dx} {\eta}=\frac{d\dot{x}} {\eta^{(1)}}$, we get
\begin{equation}
\frac{dt} {x}=\frac{dx} {-x^3}=\frac{d\dot{x}} {-(3\dot{x}x^2+\dot{x}^2)}.\label{invar}
\end{equation}
Integrating the characteristic equation (\ref{invar}) we find the invariants $u$ and $v$ are of the form
\begin{equation}
u=t-\frac{1} {x},~~v=\frac{x(\dot{x}+x^2)} {\dot{x}}.\label{vvv}
\end{equation}
The second-order invariant can be found from the relation $w=\frac{dv} {du}$ (see Eq.(\ref{cite1})). Substituting Eq. (\ref{vvv}) and their derivatives in (\ref{cite1}) and simplifying the resultant equation we arrive at
\begin{equation}
\frac{dv} {du}=x^2(1+2\frac{x^2} {\dot{x}}+\frac{x^4} {\dot{x}^2})=v^2.\label{ebt}
\end{equation}
Integrating equation (\ref{ebt}) we find $v=-\frac{1} {I_1+u}$, where $I_1$ is an integration constant. Substituting the expressions $u$ and $v$ in this solution and rewriting the resultant equation for $\dot{x}$, we end up with 
\begin{eqnarray}
\dot{x}-\frac{x-I_1x^2-tx^2} {I_1+t}=0.\label{first}
\end{eqnarray} 
Integrating Eq.(\ref{first}) we obtain the general solution of the MEE equation in the following form
\begin{equation}
x(t)=\frac{2(I_1+t)} {I_2+2I_1t+t^2},\label{lie_soln}
\end{equation}
where $I_2$ is the integration constant. The solution exactly coincides with the one found by other methods \cite{chand1,bhu1}.

\subsubsection{Linearization}

One can also identify a linearizing transformation from the Lie point symmetries if the given equation admits $sl(3,R)$ algebra \cite{new_leach}. The method of finding linearizing transformation for the modified Emden equation was discussed in detail by Mahomed and Leach \cite{mahomed1} and later by others \cite{duarte,chand1}. The underlying idea is the following. One has to choose two vector fields appropriately in such a way that they should constitute a two-dimensional algebra in the real plane \cite{19} and transform them into the canonical form $\frac{\partial}{\partial \tilde{x}}$ and $\tilde{t}\frac{\partial}{\partial \tilde{x}}$, where $\tilde{t}$ and $\tilde{x}$ are the new independent and dependent variables. For the MEE one can generate this subalgebra by considering the vector fields $V_8$ and $V_9 = V_7 -2V_6$.  To transform them into the canonical forms one should introduce the transformations \cite{nuci}
\begin{eqnarray}
 \tilde{t}=\frac{tx-1}{t(tx-2)},~~\tilde{x}=-\frac{x}{2t(tx-2)},\label{ght11}
\end{eqnarray}
where $\tilde{t}$ and $\tilde{x}$ are the new independent and dependent variables respectively. In these new variables, $(\tilde{t},\tilde{x})$, Eq.(\ref{main}) becomes the free particle equation $\frac{d^2\tilde{x}}{d\tilde{t}^2}=0$. From the solution of the free particle solution one can derive the general solution of  MEE (\ref{main}) with the help of (\ref{ght11}). The solution coincides with the one given in Eq.(\ref{lie_soln}).

\subsubsection{Lagrangian from Lie point symmetries}
Another interesting application of Lie point symmetries is that one can explore Lagrangians for the given second-order differential equation through the Jacobi last multiplier \cite{nuci}. The inverse of the determinant $\Delta$ \cite{nuci1},
 \begin{equation}
\Delta = \left| \begin{array}{ccc}
1 & \dot{x} & \ddot{x} \\
\xi_1 & \eta_1 & \eta_{1}^{(1)} \\
\xi_2 & \eta_2 & \eta_{2}^{(1)}  \end{array}\right|,~~~~M=\frac{1} {\Delta},\label{delta}
\end{equation}
where $(\xi_1,\eta_1)$ and $(\xi_2,\eta_2)$  are two sets of Lie point symmetries of the second-order ODE (\ref{main1}) and $\eta_{1}^{(1)}$ and $\eta_{2}^{(1)}$ are their corresponding first prolongations, gives the Jacobi last multiplier for the given equation. The determinant establishes the connection between the multiplier and Lie point symmetries. The Jacobi last multiplier $M$ is related to the Lagrangian $L$ through the relation \cite{nuci1}
\begin{equation}  
M=\frac{\partial^2 L}{\partial \dot{x}^2}.\label{met21}
\end{equation}
Once the multiplier is known, the Lagrangian $L$ can be derived by integrating the expression (\ref{met21}) two times with respect to $\dot{x}$. 

To obtain the Jacobi last multiplier $M$ for the MEE, we choose the vector fields $V_3$ and $V_1$. Evaluating the determinant (\ref{delta}) with these two vector fields, we find
\begin{equation}
\Delta =\left|
\begin{array}{ccc}
 1 & \dot{x} & -3x\dot{x}-x^3 \\
x & -x^3 & -\dot{x}(\dot{x}+3x^2) \\
1 & 0 & 0
\end{array}\right|=-(x^2+\dot{x})^3
\end{equation}
so that
\begin{equation}
M=-\frac{1} {(x^2+\dot{x})^3}\label{mmm1}.
\end{equation}
We can obtain the Lagrangian by integrating the expression (\ref{mmm1}) twice with respect to $\dot{x}$. Doing so, we find
\begin{eqnarray}
L&=&-\frac{1} {2(\dot{x}+x^2)}+f_1(t,x)\dot{x}+f_2(t,x),\label{lagra11}
\end{eqnarray}
where $f_1$ and $f_2$ are two arbitrary functions. The Lagrangian (\ref{lagra11}) leads to the equation of motion (\ref{main}) with $\frac{\partial f_1} {\partial t}=\frac{\partial f_2} {\partial x}$. One can also find more Lagrangains for Eq.(\ref{main1}) by considering other vector fields given in (\ref{vf8}).

\section{Noether symmetries}
\label{noeth}
In the previous section, we have discussed the invariance of the equation of motion under the one parameter Lie group of infinitesimal transformations (\ref{asjkm}). If the given second-order equation has a variational structure then one can also determine the symmetries which leave the action integral invariant. Such symmetries are called variational symmetries. Variational symmetries are important since they provide conservation laws via Noether's theorem \cite{noether}.  In the following, we recall the method of finding variational symmetries \cite{olv,blu}.

Let us consider a second-order dynamical system described by a Lagrangian $L(t,x,\dot{x})$ and
the action integral associated with the Lagrangian be
\begin{eqnarray}            
S=\int L(t,x,\dot{x}) dt.
\label {gme08}
\end{eqnarray}
Noether's theorem states that whenever the action integral is invariant under
the one-parameter group of infinitesimal transformations (\ref{asjkm}) then the solution of Euler's equation admits the
conserved quantity \cite{Gelfand:1963,Lutzky:1978}
\begin{eqnarray}            
I=(\xi\dot{x}-\eta)\frac{\partial{L}}{\partial{\dot{x}}}-\xi L+f,
\label {gme010}
\end{eqnarray}
where $f$ is a function of $t$ and $x$. The functions $\xi,\;\eta$ and $f$ can
be determined from the equation 
\begin{eqnarray}            
E\{L\}=\xi\frac{\partial{L}}{\partial{t}}+\eta\frac{\partial{L}}{\partial{x}}
+(\dot{\eta}-\dot{x}\dot{\xi})\frac{\partial{L}}{\partial{\dot{x}}},
\label {gme011}
\end{eqnarray}
where over dot denotes differentiation with respect to time and 
\begin{eqnarray}            
E\{L\}=-\dot{\xi}L+\dot{f}.
\label {gme012}
\end{eqnarray}
Equation (\ref{gme011}) can be derived by differentiating the equation
(\ref{gme010}) and simplifying the expression in the resultant equation. Solving equation (\ref{gme011}) one can obtain explicit expressions for the functions $\xi,\;\eta$ and $f$. Substituting these expressions back in (\ref{gme010}) one can get the associated integral of motion.

\subsection{Example: modified Emden equation}
In this sub-section, we illustrate the method of finding Noether symmetries and their associated conserved quantities for the MEE (\ref{main}) which has a nonstandard Lagrangian of the form (see Eq.(\ref{lagra11}))
\begin{eqnarray}            
L=\frac{1}{3(\dot{x}+x^2)},\label{lagra}
\label {kps03}
\end{eqnarray}
where we have chosen the arbitrary functions $f_1$ and $f_2$ to be zero for simplicity. Substituting the Lagrangian (\ref{lagra}) and its derivatives in (\ref{gme011}), we get
\begin{eqnarray}      
 \eta\bigg(-\frac{1} {3(\dot{x}+x^2)^2}\bigg)+
(\eta_t+\dot{x}\eta_x-\dot{x}(\xi_t+\dot{x}\xi_x))\bigg(-\frac{2x} {3(\dot{x}+x^2)^2}\bigg)\nonumber\\
=-(\xi_t+\dot{x}\xi_x)\bigg(\frac{1} {3(\dot{x}+x^2)^2}\bigg)
+f_t+\dot{x}f_x.
\label {gme014}
\end{eqnarray}

Equating the coefficients of various powers of $\dot{x}$ to zero and solving the
resultant equations, we find
\begin{eqnarray}            
\eta&=&12D-6(C+4Dt)x+3(B+3t(C+2Dt))x^2\nonumber\\
&&-\frac{3} {2}(A+t(2B+3Ct+4Dt^2))x^3,\nonumber\\           
\xi&=&E-3t(C+2Dt)+\frac{3}{2}(A+t(2B+3Ct+4Dt^2))x,\nonumber\\            
f&=&At+Bt^2+Ct^3+Dt^4,
\label {kps06}
\end{eqnarray}
where $A, B, C$, $D$ and $E$ are real arbitrary constants. The associated vector fields are
\begin{eqnarray}            
\hspace{-1cm}&&X_1=x\frac{\partial}{\partial t}
-x^3\frac{\partial}{\partial x},~~ X_2=xt\frac{\partial}{\partial t}   
	+\bigg(x^2-tx^3\bigg)\frac{\partial}{\partial x},\nonumber\\   
\hspace{-1cm}&&X_3=\bigg(t-\frac{3t^2x} {2}\bigg)\frac{\partial}{\partial t}
+\bigg(2x-3tx^2+\frac{3t^2x^3} {2}\bigg)\frac{\partial}{\partial x},\nonumber\\
\hspace{-1cm}&&X_4=\bigg(\frac{t^3x} {2}-\frac{t^2} {2}\bigg)\frac{\partial}{\partial t}
+\bigg(1-2tx +\frac{3t^2x^2} {2}-\frac{t^3x^3} {2}\bigg)\frac{\partial}{\partial x},~X_5=\frac{\partial}{\partial t}.    
\label {sgme016}
\end{eqnarray}

The Noether's symmetries are sub-set of Lie point symmetries. In the above, while the vector fields $X_1, X_2$ and $X_5$ exactly match with the Lie vector fields $V_3,V_4$ and $V_1$ (see Eq. (\ref{vf8})), the remaining two Noether vector fields $X_3$ and $X_4$ can be expressed as linear combinations of other Lie point symmetries, that is $X_3=V_2+2V_5$ and $X_4=V_7-2V_6$.

Substituting each one of the vector fields separately into (\ref{gme010}) we obtain the associated integrals of motions. They turned out to be
\begin{eqnarray}            
\hspace{-1.2cm}&&I_1= t-\frac{x}{x^2+\dot{x}},~~I_2= \frac{(-x+tx^2+t\dot{x})^2}{(x^2+\dot{x})^2},
\nonumber\\       
\hspace{-1.2cm}&&I_3=\frac{-9t^2x^3+3t^3x^4-3x(2+3t^2\dot{x})+2tx^2(6+3t^2\dot{x})
+9t\dot{x}(6+3t^2\dot{x})}{(x^2+\dot{x})^2},
\nonumber\\           
\hspace{-1.2cm}&&I_4=\frac{3(2-2tx+t^2x^2+t^2\dot{x})}{(x^2+\dot{x})},
\nonumber\\ 
\hspace{-1.2cm}&&I_5=\frac{2\dot{x}+x^2}{3(x^2+\dot{x})^2},\qquad\qquad \qquad \qquad\frac{dI_i}{dt}=0,~i=1,2,3,4,5.
\label {kps11}
\end{eqnarray}

One can easily verify that out of the five integrals two of them are independent and the remaining three can be expressed in terms of these two integrals, that is $I_2=I_1^2,~~ I_3=I_1I_4$ and $I_5=\frac{1}{9}(I_4-3I_1^2)$. We can construct a general solution of (\ref{main}) with the help of the two independent integrals $I_1$ and $I_4$. The underlying solution coincides with (\ref{lie_soln}) after rescaling.

\section{Contact symmetries}
\label{comta}
In the previous two cases, Lie point symmetries and Noether symmetries, we have considered the functions $\xi$ and $\eta$ to be functions of $t$ and $x$ only. One may relax this condition by allowing the functions $\xi$ and $\eta$ to depend on $\dot{x}$ besides $t$ and $x$. This generalization was considered Sophus Lie himself \cite{19} and later by several authors \cite{ci4,ci5,sir_old1,sir_old2}. This velocity dependent infinitesimal transformations are called contact transformations and the functions $\xi$ and $\eta$ are called contact symmetries. The contact symmetries for the harmonic and damped harmonic oscillators were worked out explicitly by Schwarz and Cerver$\acute{o}$ and Villarroel respectively \cite{ci4,ci5}. Several nonlinear second-order ODEs do not admit Lie point symmetries but they were proved to be integrable by other methods. To demonstrate the integrability of these nonlinear ODEs in Lie's sense one should consider velocity dependent transformations. In the following, we give a brief account of the theory and illustrate the underlying ideas by considering MEE as an example.
\subsection{Method of Lie}
Let a one-parameter group of contact transformation be given by \cite{ci4}
\begin{eqnarray}
\hspace{-1cm}T=t+\varepsilon \,\xi(t,x,\dot{x}),~X=x+\varepsilon \,\eta(t,x,\dot{x}),~\dot{X}=\dot{x}+\varepsilon \, \eta^{(1)}(t,x,\dot{x}) \quad \epsilon \ll 1.\label{asm}
\end{eqnarray}
The functions $\xi$ and $\eta$ determine an infinitesimal contact transformation if it is possible to write them in the form \cite{ci7}
\begin{eqnarray}
\xi(t,x,\dot{x})=-\frac{\partial W} {\partial \dot{x}},~~\eta(t,x,\dot{x})=W-\dot{x}\frac{\partial W} {\partial \dot{x}},~~\eta^{(1)}=\frac{\partial W} {\partial t}+\dot{x}\frac{\partial W} {\partial x},
\end{eqnarray} where the characteristic function $W(t,x,\dot{x})$ is an arbitrary function of its arguments. If $W$ is linear in $\dot{x}$ the corresponding contact transformation is an extended point transformation and it holds that $W(t,x,\dot{x})=\eta(t,x)-\dot{x}\xi(t,x)$. A second-order differential equation (\ref{main1}) is said to be invariant under the contact transformation (\ref{asm}) if $\xi \frac{\partial \phi}{\partial t}+\eta \frac{\partial \phi}{\partial x}+\eta^{(1)}\frac{\partial \phi}{ \partial \dot x}-\eta^{(2)} =0$ on the manifold $\ddot{x}-\phi(t,x,\dot{x})=0$ in the space of the variables $t,x,\dot{x}$ and $\ddot{x}$ \cite{ci4,ci5}. Here $\eta^{(1)}$ and $\eta^{(2)}$ are the first and second prolongations with $\eta^{(1)}=\dot{\eta}-\dot{x}\dot{\xi}$ and $\eta^{(2)}=\dot{\eta}^{(1)}-\ddot{x}\dot{\xi}$. The invariance condition provides the following linear partial differential equation for the characteristic function $W$:
\begin{eqnarray}
\hspace{-1.4cm}&&\frac{\partial W}{\partial \dot{x}} \frac{\partial \phi}{\partial t}  +\left(\dot{x} \frac{\partial W}{\partial \dot{x}}-W\right)\frac{\partial \phi}{\partial x} - \left(\frac{\partial W}{\partial t} +\dot{x} \frac{\partial W}{\partial x}\right) \frac{\partial \phi}{\partial \dot{x}} +
\bigg(\phi^2 \frac{\partial ^2 W}{\partial \dot{x}^2}+2 \phi \frac{\partial^2 W }{\partial t \partial \dot{x}}\nonumber \\ \hspace{-1.3cm}&&\qquad\qquad \quad+2 \phi \dot{x} \frac{\partial^2 W}{\partial x \partial \dot{x}}+\phi \frac{\partial W}{\partial x}+\frac{\partial ^2W}{\partial t^2}+2 \dot{x} \frac{\partial^2 W}{\partial t\partial x}+\dot{x}^2 \frac{\partial ^2W}{\partial x^2}\bigg)=0.\label{cintchar}
\end{eqnarray}
Integrating Eq.(\ref{cintchar}) one can get the characteristics $W$. From $W$ one can recover the contact symmetries $\xi$ and $\eta$. One can also recover the necessary independent integrals from the characteristic function (see for example, Ref.\cite{ci4}).
\subsection{Example: modified Emden equation}
To determine the contact symmetries of MEE, we have to determine the characteristic function (\ref{cintchar}), by solving the first-order partial differential equation
\begin{eqnarray}
\hspace{-1.4cm}&&-(3\dot{x}+3x^2)\left(\dot{x} \frac{\partial W}{\partial \dot{x}}-W\right) +3x \left(\frac{\partial W}{\partial t} +\dot{x} \frac{\partial W}{\partial x}\right)  +\bigg((3x\dot{x}+x^3)^2 \frac{\partial ^2 W}{\partial \dot{x}^2}\nonumber \\ \hspace{-1.4cm}&&\quad\quad \quad-2 (3x\dot{x}+x^3)\frac{\partial^2 W}{\partial t \partial \dot{x}}-2 (3x\dot{x}+x^3) \dot{x} \frac{\partial^2 W}{\partial x \partial \dot{x}}-(3x\dot{x}+x^3) \frac{\partial W}{\partial x}\nonumber \\ \hspace{-1.7cm}&&\qquad\qquad \qquad\qquad\qquad\qquad\qquad+\frac{\partial ^2W}{\partial t^2}+2 \dot{x}\frac{\partial^2 W}{\partial t \partial x}+\dot{x}^2 \frac{\partial ^2W}{\partial x^2}\bigg)=0.\label{cintchar1}
\end{eqnarray}

One may find the general solution of the above linear partial differential equation by employing the well known methods for solving linear partial differential equations. In general $W$ depends upon arbitrary functions and the contact Lie group has an infinite number of parameters. 

For the sake of illustration, in the following, we present two particular solutions of Eq.(\ref{cintchar1}):
\begin{eqnarray} 
W_1=\frac{x^2 (1-t x)}{x^2+\dot{x}}-t^2 \dot{x},~~W_2=-\frac{x \dot{x}}{\sqrt{x^2+2 \dot{x}}}-\frac{x \left(x^2+\dot{x}\right)}{\sqrt{x^2+2 \dot{x}}}.
\end{eqnarray}
The infinitesimal vector fields read
\begin{eqnarray}
\hspace{-0.5cm}\Omega_1=t^2\frac{\partial}{\partial t}+\frac{x^2(1-tx)} {\dot{x}+x^2}\frac{\partial}{\partial x},~~\Omega_2=\frac{x}{\sqrt{2\dot{x}+x^2}}\frac{\partial}{\partial t}-x\frac{(\dot{x}+x^2)} {\sqrt{2\dot{x}+x^2}}\frac{\partial}{\partial x}.\label{kinl}
\end{eqnarray}
As one can see from (\ref{kinl}) the infinitesimals $\xi$ and $\eta$ depend on the velocity terms also. One can derive the general solution from each one of the contact symmetries by solving the characteristic equation associated with the vector field. We demonstrate this procedure in Sec. \ref{con_sub}.

\subsection{Method of Gladwin Pradeep et.al}
\label{con_sub}
In a recent paper Gladwin Pradeep et.al proposed a new method of finding contact symmetries for a class of equations \cite{cont}. Their method involves two steps. In the first step, one has to find a linearizing contact transformation. In the second step, with the help of contact transformation, one proceeds to construct contact symmetries for the given equation. Once the contact symmetries are determined the order reduction procedure can be employed to derive the general solution of the given differential equation. In the following, we recall this procedure with MEE as an example.

\vspace{0.2cm}
{\it Step 1: Linearizing contact transformation} 
\vspace{0.2cm}

The MEE (\ref{main}) can be linearized to the free particle equation $\frac{d^2u} {dt^2}=0$ by the contact transformation (for more details one may refer, Ref.\cite{cont})
\begin{eqnarray}
x=\frac{2u\dot{u}}{1+u^2},\qquad\dot{x}=\frac{2\dot{u}^2(1-u^2)}{(1+u^2)^2}.\label{contscx}
\end{eqnarray}
One may also invert the above transformation (\ref{contscx}) and obtain
\begin{eqnarray}
u=\frac{x}{\sqrt{2\dot{x}+x^2}},~~
\dot{u}=\frac{(\dot{x}+x^2)}{\sqrt{2\dot{x}+x^2}}.\label{conct2}
\end{eqnarray}

\vspace{0.2cm}
{\it Step 2: Contact symmetries}
\vspace{0.2cm}

 Let us designate the symmetry vector field and its first prolongation of the nonlinear ODE (\ref{main}) respectively be of the form
$
\Omega=\lambda \frac{\partial}{\partial t}
+\mu \frac{\partial}{\partial x},
$
and
\begin{eqnarray}            
\Omega^{1}=\lambda \frac{\partial}{\partial t}
+\mu \frac{\partial}{\partial x}+(\dot{\mu}
-\dot{x}\dot{\lambda})\frac{\partial}{\partial \dot{x}},
\label {sym04}
\end{eqnarray}
where $\lambda$ and $\mu$ are the infinitesimals associated with the variables $t$ and $x$, respectively.  

Let the symmetry vector field associated with the linear ODE, $\frac{d^2u} {dt^2}=0$, be $\Lambda=\xi \frac{\partial}{\partial t}+\eta \frac{\partial}{\partial u},$ and its first extension be $\Lambda^{1}=\xi \frac{\partial}{\partial t}+\eta \frac{\partial}{\partial u}+(\dot{\eta}
-\dot{u}\dot{\xi})\frac{\partial}{\partial \dot{u}}$. Using the contact transformation (\ref{conct2}) we can deduce the following differential identities, that is $\frac{\partial }{\partial u}=\frac{\partial x}{\partial u}\frac{\partial}{\partial x}+\frac{\partial \dot{x}}{\partial u}\frac{\partial}{\partial \dot{x}}$ and $\frac{\partial }{\partial \dot{u}}=\frac{\partial x}{\partial \dot{u}}\frac{\partial}{\partial x}+\frac{\partial \dot{x}}{\partial \dot{u}}\frac{\partial}{\partial \dot{x}}$. Rewriting the first prolongation $\Lambda^1$ using these relations, we find
\begin{eqnarray}
\hspace{-0.8cm}\Lambda^1=\xi\frac{\partial}{\partial t}+\left(\eta\frac{\partial x}{\partial u}+(\dot{\eta}-\dot{u}\dot{\xi})\frac{\partial x}{\partial \dot{u}}\right)\frac{\partial}{\partial x}+
\left(\eta\frac{\partial \dot{x}}{\partial u}+(\dot{\eta}-\dot{u}\dot{\xi})\frac{\partial \dot{x}}{\partial \dot{u}}\right)\frac{\partial}{\partial \dot{x}}.\label{lambda1}
\end{eqnarray}
Now comparing the vector fields (\ref{lambda1}) and (\ref{sym04}), we find
\begin{eqnarray}
\hspace{-0.7cm}\lambda=\xi,\qquad
\mu=\eta\frac{\partial x}{\partial u}+(\dot{\eta}-\dot{u}\dot{\xi})\frac{\partial x}{\partial \dot{u}}=\frac{\sqrt{2\dot{x}+x^2}}{\dot{x}+x^2}\left(\eta\dot{x}+\dot{\eta}x\right)-\dot{\xi}x.\label{arbitrary-sym}
\end{eqnarray}
The functions $\xi$ and $\eta$ are the symmetries of the free particle equation. They are given by \cite{olv,hydon}
\begin{eqnarray} 
&&\Lambda_1=\frac{\partial}{\partial t},\quad            
\Lambda_2=\frac{\partial}{\partial u},\quad 
\Lambda_3=t \frac{\partial}{\partial u},\quad 
\Lambda_4=u\frac{\partial}{\partial u},\quad
\Lambda_5=u \frac{\partial}{\partial t},\nonumber\\ 
&&\Lambda_6=t \frac{\partial}{\partial t}
,\quad 
\Lambda_7=t^2 \frac{\partial}{\partial t}
+tu \frac{\partial}{\partial u},\quad 
\Lambda_8=tu \frac{\partial}{\partial t}
+u^2 \frac{\partial}{\partial u}.
\label {sym11}
\end{eqnarray}
Substituting the above symmetry generators $\Lambda_i$'s, $i=2,\ldots,8,$ in Eq. (\ref{arbitrary-sym}), we can determine the function $\mu$. Substituting $\lambda=\xi$ and $\mu$ in the vector field $\Omega=\xi \frac{\partial}{\partial t}+\mu \frac{\partial}{\partial x}$, we arrive at the following contact symmetry generators of Eq. (\ref{main}),
\begin{eqnarray}            
&&\Omega_1=\frac{\partial}{\partial t},~~\Omega_2=\frac{\sqrt{2\dot{x}+x^2}}{\dot{x}+x^2}\dot{x}\frac{\partial}{\partial x},~~\Omega_3=\frac{\sqrt{2\dot{x}+x^2}}{\dot{x}+x^2}(t\dot{x}+x)\frac{\partial}{\partial x},\nonumber\\
&&\Omega_4=\frac{x(2\dot{x}+x^2)}{(\dot{x}+x^2)}\frac{\partial}{\partial x},~~\Omega_5=\frac{x}{\sqrt{2\dot{x}+x^2}}\frac{\partial}{\partial t}-x\frac{(\dot{x}+x^2)} {\sqrt{2\dot{x}+x^2}}\frac{\partial}{\partial x},\nonumber\\&& \Omega_6=t\frac{\partial}{\partial t}-x\frac{\partial} {\partial x},
~~ \Omega_7=t^2\frac{\partial}{\partial t}+\frac{x^2(1-tx)} {\dot{x}+x^2}\frac{\partial}{\partial x},\nonumber\\
&&\Omega_8=\frac{tx}{\sqrt{(2\dot{x}+x^2)}}\frac{\partial}{\partial t}+\bigg(\frac{x^2\sqrt{(2\dot{x}+x^2)}} {\dot{x}+x^2}- \frac{tx(\dot{x}+x^2)} {\sqrt{(2\dot{x}+x^2)}}\bigg)\frac{\partial}{\partial x}.\label{dyna_symm}
\end{eqnarray}
One can unambiguously verify that all the symmetry generators are solutions of the invariance condition.

\subsubsection{General solution} 

To derive the general solution of the given equation one has to integrate the Lagrange's system associated with the contact symmetries given above. We demonstrate this procedure by considering the vector field $\Omega_4$ given in Eq.(\ref{dyna_symm}). For the other vector fields one may follow the same procedure. 

The characteristic equation associated with the vector field $\Omega_4$ turns out to be
\begin{equation}
\frac{dt}{0}=\frac{(\dot{x}+x^2)dx}{x(2\dot{x}+x^2)}=-\frac{(\dot{x}+x^2)d\dot{x}}{x^4+x^2\dot{x}-2\dot{x}^2}. \label{inpo}
\end{equation}
Integrating Eq.(\ref{inpo}) we find the invariants $u$ and $v$ are of the form $u=t$ and $v=\frac{\dot{x}}{x}+x$. In terms of these variables one can reduce the order of the equation (\ref{main}). The reduced equation turns out to be the Riccati equation,
$\frac{dv}{dt}=-v^2$, whose general solution is given by $v=\frac{1}{I_1+u},$
where $I_1$ is the integration constant. Substituting the expressions $u$ and $v$ in the above solution and rearranging the resultant expression for $\dot{x}$, we find
\begin{equation}
\dot{x}-\frac{x} {I_1+t}+x^2=0.\label{riccati-eg1}
\end{equation}
Integrating (\ref{riccati-eg1}) one can obtain the general solution of (\ref{main}). The general solution coincides with (\ref{lie_soln}) after appropriate rescaling.

\section{Symmetries and integrating factors}
\label{sec5}
In Sec.\ref{2}, we have discussed only few applications of Lie point symmetries. One can also determine integrating factors from Lie point symmetries. In fact, Lie himself found the equivalence between integrating factors and Lie point symmetries for the first-order ODEs. For second-order ODEs the equivalence has been established only recently \cite{mur1}. The reason is that unlike the first-order ODEs (which admit infinite number of symmetries) several second-order ODEs do not admit Lie point symmetries although they are integrable by quadratures. Subsequently attempts have been made to generalize the classical Lie method to yield nontrivial symmetries and integrating factors. Two such generalizations which come out in this direction are (i) adjoint symmetries method \cite{blu,blu11} and (ii) $\lambda$-symmetries approach \cite{mur3,mur4}. The adjoint symmetry method was developed  by Bluman and Anco \cite{blu1} and the $\lambda$-symmetry approach was floated by Muriel and Romero. The applicability of both the methods have been demonstrated for the equations which lack Lie point symmetries. In both the methods one can find symmetries, integrating factors and integrals associated with the given equation in an algorithmic way.

In the following, we recall briefly these two powerful methods and demonstrate the underlying ideas by considering MEE as an example.
 
\subsection{Method of Bluman and Anco}

In general, an integrating factor is a function, multiplying the ODE, which yields a first integral. If the given ODE is self-adjoint, then its integrating factors are necessarily solutions of its linearized system (\ref{met1411}) \cite{blu}. Such solutions are the symmetries of the given ODE. If a given ODE is not self-adjoint, then its integrating factors are necessarily solutions of the adjoint system of its linearized system. Such solutions are known as adjoint symmetries of the given ODE \cite{blu1}. 

Let us consider second-order ODE (\ref{main1}). The linearized ODE for Eq.(\ref{main1}) is given in (\ref{met1411}). The adjoint ODE of the linearized equation is found to be
\begin{equation}
L^*[x]w=\frac{d^2w} {dt^2}+\frac{d} {dt}(\phi_{\dot{x}}w)-\phi_xw=0.\label{ajoin1}
\end{equation}
The solutions $w=\Lambda(t,x,\dot{x})$ of the above equation (\ref{ajoin1}) holding for any $x(t)$ satisfying the given equation (\ref{main1}) are the adjoint symmetries of (\ref{main1}) \cite{blu1}. 


The adjoint symmetry of the Eq.(\ref{main1}) becomes an integrating factor of (\ref{main1}) if and only if $\Lambda(t,x,\dot{x})$ satisfies the adjoint invariance condition \cite{blu11}
\begin{eqnarray}
L^*[x]\Lambda(t,x,\dot{x})=-\Lambda_{x}(\ddot{x}-\phi)+\frac{d} {dt}(\Lambda_{\dot{x}}(\ddot{x}-\phi)). \label{comp1}
\end{eqnarray}
Now comparing the Eqs. (\ref{ajoin1}) and (\ref{comp1}) and collecting the powers of $\ddot{x}$ and constant terms, we find
\numparts
\addtocounter{equation}{-1}
\label{adjon_con}
\addtocounter{equation}{1}
\begin{eqnarray}
&&\Lambda_{t\dot{x}}+\Lambda_{x\dot{x}}\dot{x}+2\Lambda_{x}+\Lambda \phi_{\dot{x}\dot{x}}+2\phi_{\dot{x}}\Lambda_{\dot{x}}+\phi \Lambda_{\dot{x}\dot{x}}=0,\label{adjo1}\\
&&\Lambda_{tt}+2\Lambda_{tx}\dot{x}+\Lambda_{xx}\dot{x}^2+\Lambda \phi_{t\dot{x}}+\Lambda \phi_{x\dot{x}}\dot{x}+\phi_{\dot{x}}\Lambda_{t}+\phi_{\dot{x}}\Lambda_{x}\dot{x}\nonumber \\ &&-\Lambda \phi_{x}-\Lambda_{x}\phi+\phi \Lambda_{t\dot{x}}+\phi \Lambda_{x\dot{x}}\dot{x}+\Lambda_{\dot{x}}\phi_t+\Lambda_{\dot{x}}\phi_{x}\dot{x}=0.\label{adjo2}
\end{eqnarray}
\endnumparts
The solutions of (\ref{adjo2}) are called the adjoint symmetries. If these solutions also satisfy Eq.(\ref{adjo1}) then they become integrating factors for the given second-order ODE (\ref{main1}). 

The main advantage of this method is that if the given equation is of odd order or does not have variational structure, one can use this method and obtain the integrals in an algorithmic way.

\subsubsection{Example: modified Emden equation}
For the MEE equation, the linearized equation is given by
\begin{equation}
\frac{d^2Q} {dt^2}+3x\frac{dQ} {dt}+(3\dot{x}+3x^2) Q=0.
\label{met1411ad}
\end{equation}
The adjoint equation for the above linearized equation turns out that
\begin{equation}
\frac{d^2w} {dt^2}+\frac{d} {dt}(-3xw)+(3\dot{x}+3x^2)w=0.\label{adj_exam}
\end{equation}
Since Eqs. (\ref{met1411ad}) and (\ref{adj_exam}) do not coincide, the function $Q$ is not an integrating factor for the MEE. In this case the integrating factors can be determined from the adjoint symmetry condition (\ref{adj_exam}). The adjoint symmetry determining equation (\ref{adj_exam}) for the present example reads
\begin{eqnarray}
&&\hspace{-1.3cm}\Lambda_{tt}+2\Lambda_{tx}\dot{x}+\Lambda_{xx}\dot{x}^2-3\Lambda\dot{x}-3x\Lambda_{t}-3 x\Lambda_{x}\dot{x}+(3\dot{x}+3x^2)\Lambda \nonumber \\ &&\hspace{-1.8cm}+(3 x\dot{x}+x^3)\Lambda_{x}-(3 x \dot{x}+x^3)\Lambda_{\dot{xt}}-(3x\dot{x}+x^3)\dot{x}\Lambda_{x\dot{x}}-(3 \dot{x}+3x^2)\Lambda_{\dot{x}}\dot{x}=0.\label{adjo3}
\end{eqnarray} 
Two particular solutions of (\ref{adjo3}) are given by
\begin{equation}
\Lambda_1= \frac{x} {(\dot{x}+x^2)},~~\Lambda_{2}=\frac{t(-2+tx)} {2(t\dot{x}-x+tx^2)^2}.
\label{adm112}
\end{equation}
 
These two adjoint symmetries, $\Lambda_1$ and $\Lambda_2$, also satisfy the Eq.(\ref{adjo1}). So they become integrating factors for the Eq.(\ref{main}). Multiplying the given equation by each one of these integrating factors and rewriting the resultant expression as a perfect derivative and integrating them we obtain two integrals $I_1$ and $I_2$ which are of the form,
\begin{eqnarray}
I_1&=&-\frac{(\dot{x}t-x+tx^2)} {\dot{x}+x^2},~~I_2=-\frac{(2+\dot{x}t^2-2tx+t^2x^2)} {2(\dot{x}t-x+tx^2)}.\label{lam117}
\end{eqnarray}
From these integrals, $I_1$ and $I_2$, the general solution can be derived. The resultant expression coincides with Eq.(\ref{lie_soln}) after rescaling.

\subsection{Method of Muriel and Romero}
As we mentioned at few places earlier, many second-order nonlinear dynamical systems often lack Lie point symmetries but proved to be integrable by other methods. To overcome this problem, efforts have been made to generalize the classical Lie algorithm and obtain integrals and general solution of these nonlinear ODEs. One such generalization is the $\lambda$-symmetries approach. This approach was developed by Muriel and Romero \cite{mur1}. These symmetries are neither Lie point nor Lie-B$\ddot{a}$cklund symmetries and are called $\lambda$-symmetries since they are vector fields that depend upon a function $\lambda$. If we choose this arbitrary function as null we obtain the classical Lie point symmetries. The method of finding $\lambda$-symmetries for a second-order ODE has been discussed in depth by Muriel and Romero and the advantage of finding such symmetries has also been demonstrated by them \cite{mur3}. The authors have also developed an algorithm to determine integrating factors and integrals from $\lambda$-symmetries for the second-order ODEs \cite{mur4}. 

Consider a second-order ODE (\ref{main1}). Let $\tilde{V}=\xi(t,x)\frac{\partial}{\partial t}+\eta(t,x)\frac{\partial}{\partial x}$ be a $\lambda$-symmetry of the given ODE for some function $\lambda=\lambda(t,x,\dot{x})$. The invariance of the given ODE under $\lambda$-symmetry vector field is given by \cite{mur1}
\begin{equation}
\xi\phi_t+\eta\phi_x+\eta^{[\lambda,(1)]}\phi_{\dot{x}}-\eta^{[\lambda,(2)]}=0,
\label{beq1}
\end{equation}
where $\eta^{[\lambda,(1)]}$ and $\eta^{[\lambda,(2)]}$ are the first and second $\lambda$- prolongations. They are given by
\begin{eqnarray}
\hspace{-1cm}\eta^{[\lambda,(1)]}&=&(D_t+\lambda)\eta^{[\lambda,(0)]}(t,x)-(D_t+\lambda)(\xi(t,x))\dot{x}=\eta^{(1)}+\lambda(\eta-\dot{x}\xi),\label{beq2}\nonumber \\
\hspace{-1cm}\eta^{[\lambda,(2)]}&=&(D_t+\lambda)\eta^{[\lambda,(1)]}(t,\dot{x})-(D_t+\lambda)(\xi(t,x))\ddot{x}=\eta^{(2)}+f(\lambda),
\end{eqnarray}
where $f(\lambda)$ is given by $D[\lambda](\eta-\xi \dot{x})+2\lambda(\eta^{(1)}-\xi\ddot{x})+\lambda^2(\eta-\xi \dot{x})$ and $\eta^{(0)}=\eta(t,x)$, and $D$ is the total differential operator $(D=\frac{\partial}{\partial t}+\dot{x}\frac{\partial}{\partial x}+\phi \frac{\partial}{\partial \dot{x}})$.  In the above prolongation formula if we put $\lambda=0$, we end up at standard Lie prolongation expressions. Solving the invariance condition (\ref{beq1}) we can determine the functions $\xi$, $\eta$ and $\lambda$ for the given equation. We note here that three unknowns $\xi$, $\eta$ and $\lambda$ have to be determined from the invariance condition (\ref{beq1}). The procedure is as follows.

Let us suppose that the second-order equation (\ref{main1}) has Lie point symmetries. In that case, the $\lambda$-function can be determined in a more simple way  without solving the invariance condition (\ref{beq1}).

If $V$ is a Lie point symmetry of (\ref{main1}) and $Q=\eta-\dot{x}\xi$ is its characteristics then $v=\frac{\partial} {\partial x}$ is a $\lambda$-symmetry of (\ref{main1}) for $\lambda=\frac{D[Q]} {Q}$. The $\lambda$-symmetry satisfies the invariance condition \cite{mur3} 
\begin{equation}
\phi_x+\lambda \phi_{\dot{x}}=D[\lambda]+\lambda^2.\label{lamd}
\end{equation}

 Once the $\lambda$-symmetry is determined, we can obtain the first integrals in two different ways.  In the first way, we can calculate the integral directly from the $\lambda$-symmetry using the four step algorithm given below. In the second way, we can find the integrating factor $\mu$ from $\lambda$-symmetry. With the help of integrating factors and $\lambda$-symmetries we can obtain the first integrals by integrating the system of Eqs.(\ref{imueq}). In the following, we enumerate both the procedures.

\vspace{0.2cm}
{\it (A) Method of finding the first integral directly from $\lambda$-symmetry \cite{mur3}}
\vspace{0.2cm}

The method of finding integral
directly from $\lambda$-symmetry is as follows:
\begin{enumerate}[(i)]
\item
Find a first integral $w(t,x,\dot x)$ of $v^{[\lambda,(1)]}$, that is a particular solution of the equation
$w_x+\lambda w_{\dot{x}}=0,
$
where subscripts denote partial derivative with respect to that variable and $v^{[\lambda,(1)]}$ is the first-order $\lambda$-prolongation of the vector field $v$.
\item
Evaluate $D[w]$ and express $D[w]$ in terms of $(t,w)$ as $D[w]=F(t,w)$.
\item
Find a first integral G of $\partial_t+F(t,w)\partial_w$.
\item
Evaluate $I(t,x,\dot{x})=G(t,w(t,x,\dot{x}))$.
\end{enumerate}

\vspace{0.2cm}
{\it (B) Method of finding integrating factors from $\lambda $ \cite{mur3}}
\vspace{0.2cm}

{\it 
If $V$ is a Lie point symmetry of (\ref{main1}) and $Q=\eta-\dot{x}\xi$ is its characteristics, then
$v=\partial_{x}$ is a $\lambda$-symmetry of (\ref{main1}) for $\lambda=D[Q]/Q$ and any solution of the first-order linear system
\begin{eqnarray}
D[\mu]+(\phi_{\dot{x}}-\frac{D[Q]}{Q})\mu = 0, \;\;\;
\mu_x+(\frac{D[Q]}{Q}\mu)_{{\dot{x}}} = 0,
\label{musaa}
\end{eqnarray}
is an integrating factor of (\ref{main1}). Here $D$ represents the total derivative operator and it is given by $\frac{\partial}{\partial t}+\dot{x}\frac{\partial}{\partial x}+\phi \frac{\partial}{\partial \dot{x}}$.}

Solving the system of equations (\ref{musaa}) one can get $\mu$.  Once the integrating factor $\mu$ is known then a first integral $I$ such that $I_{\dot{x}}=\mu$ can be
found by solving the system of equations
\begin{eqnarray}
I_t=\mu(\lambda \dot{x}-\phi),\;\; I_x=-\lambda \mu,\;\; I_{\dot{x}}=\mu.
\label{imueq}
\end{eqnarray}
From the first integrals, we can write the general solution of the given equation.

\subsubsection{Example: modified Emden equation}
\label{lamb_sec}
Since the second-order ODE under investigation admits Lie point symmetries one can derive the $\lambda$-symmetries directly from Lie point symmetries \cite{bhu1} through the relation  $\lambda=\frac{D[Q]}{Q}$.

To start with, we consider the vector field $V_3$. In this case, we have $\xi = x$ and $\eta = -x^3$, the $Q$ function turns out to be $Q=\eta-\xi \dot{x}$ = $-x(\dot{x}+x^2)$.  Using the relation $\lambda=\frac{D[Q]}{Q}$ we can fix $\lambda _3 = \frac{\dot{x}}{x}-x$. In a similar way one can fix the $\lambda$-symmetries for the remaining vector fields. The resultant expressions are given in the following Table. One can verify that the functions $\xi$, $\eta$ and $\lambda$ satisfy the invariance condition (\ref{beq1}).  
{\tiny
\begin{table}[h]
\begin{center}
\begin{tabular}{|c|c|c|}
\hline
Vector & Q & $\lambda$-symmetries  \\
\hline
$V_1$& $-\dot{x}$  &$-(3 x +\frac{x^3}{\dot x})$ \\
\hline
$V_2$  & $\frac{1} {2}t(-2+tx) (\dot{x}+x^2)$ & $\frac{2-\dot{x}t^2-4 t x+t^2 x^2}{t(2-t x)}$ \\
\hline
$V_3$ & $-x(\dot{x}+x^2)$  &$\frac{\dot{x}} {x}-x$ \\
\hline
$V_4$ & $x(-\dot{x}t+x-tx^2)$  & $\frac{\dot{x}} {x}-x$ \\
\hline
$V_5$ & $\frac{x} {2}(2+\dot{x}t^2-2 t x+t^2 x^2)$  &$\frac{\dot{x}} {x}-x$ \\
\hline
$V_6$ & $\frac{t} {2}(-2+tx)(\dot{x}t-x+tx^2)$  & $(\frac{2-\dot{x}t^2-4 t x+t^2 x^2}{t(2-t x)})$ \\
\hline
$V_7$ & $\frac{1} {2}(2-3t^2x^2+t^3x^3-3\dot{x}t^2+\dot{x}t^3x$  &$-\frac{t(-\dot{x}^2t^2+\dot x(6-6tx)+x^2(6-6tx+t^2x^2)} {2-3t^2x^2+t^3x^3+\dot{x}t^2(-3+tx)}$ \\
\hline
$V_8$ & $\frac{-t} {4}(-2+tx) (2+\dot{x}t^2-2 tx+t^2x^2$   &$\frac{2-\dot{x}t^2-4 t x+t^2 x^2}{t(2-t x)}$  \\
\hline
\end{tabular}
\caption{$\lambda$- symmetries for eight the vector fields admitted by Eq.(\ref{main})}
\end{center}
\end{table}}

Since we are dealing with a second-order ODE two different $\lambda$ functions are sufficient to generate two independent first integrals and hence the general solution. In the following, we consider the vector fields $V_3$ and $V_6$ and their $\lambda$-symmetries and demonstrate the method of finding their associated integrals.

 \vspace{0.3cm}
{\it (i) First integrals directly from $\lambda$-symmetry}
\vspace{0.3cm}


Substituting $ \lambda _3 = \frac{\dot x}{x}-x$ in the equation $w_x+\lambda w_{\dot{x}}=0,
$ one gets $w_x+\left(\frac{\dot x}{x}-x\right)w_{\dot{x}}=0$. This first-order PDE admits an integral $w(t,x,\dot{x})$ of the form $w(t,x,\dot{x}) = \frac{\dot x}{x}+x$ (first step). The total differential of this function can be expressed in terms of $w$ itself, that is $D[w] =-w^2= F(t,w)$  (second step). Next, one has to find an integral associated with the first-order partial differential equation $\frac{\partial G}{\partial t}-w^2\frac{\partial G}{\partial w}=0$. A particular solution of this first-order partial differential equation can be given as $G=t-\frac{1}{w}$ (third step). Finally, one has to express $G(t,w)$ in terms of $(t,x,\dot{x})$.  Doing so, we find that the integral turns out to be (fourth step)
\begin{equation}
\hat{I}_1=t-\frac{x}{\dot{x}+x^2},~~~~~\qquad\frac{d\hat{I}_1}{dt}=0.\label{i11}
\end{equation}

Next we consider the function $\lambda _6$. Following the steps given above, we find the integral associated with $\lambda_6$ which turns out to be
\begin{equation}
\hat{I}_2=\frac{\dot{x}t-x+tx^2} {2+\dot{x}t^2-2tx+t^2x^2}
\label{i12}
\end{equation}
with $\frac{d\hat{I}_2}{dt}=0$. Using these two integrals, $\hat{I}_1$ and $\hat{I}_2$, one can construct the general solution of Eq.(\ref{main}).  The resultant expression coincides with the earlier expression (see Eq.(\ref{lie_soln})) after rescaling.
\vspace{0.5cm}

{\it (ii) Integrating factors from $\lambda$-symmetries}
\vspace{0.3cm}

In the following, we discuss the second route of obtaining the integral from $\lambda$-symmetry.  We solve the system of equations (\ref{musaa}) in the following way. We first consider the second equation in (\ref{musaa}) and obtain a solution for $\mu$.  We then check whether the obtained expression satisfies the first equation or not.  If it satisfies then we treat it as a compatible solution.

We again consider the Lie point symmetries $V_3$ and $V_6$ and discuss the method of deriving integrating factors for these functions. To determine the integrating factor associated with $\lambda _3$ we first solve the second equation in (\ref{musaa}), that is $\mu_x+(\frac{\dot{x}}{x}-x)\mu_{\dot{x}}+\frac{1}{x}\mu=0$. A particular solution is $\mu_1 = -\frac{x}{(\dot{x}+x^2)^2}$. This solution also satisfies the first equation in (\ref{musaa}). To determine the integral we substitute $\mu_1$ and $\lambda _3$ in (\ref{imueq}) and obtain the following set of equations for the unknown $I$, namely
\begin{eqnarray}
I_t=1,\;\; I_x=-\frac{(\dot{x}-x^2)}{(\dot{x}+x^2)^2},\;\;
I_{\dot{x}}=\frac{x}{(\dot{x}+x^2)^2}.
\label{mu2eq333}
\end{eqnarray}
The integral which comes out by integrating the system of equations (\ref{mu2eq333}), that is $I_1=t-\frac{x}{x^2+\dot x}$, coincides exactly with the one found earlier.

Let us now consider the function $\lambda_6$. Substituting the function $\lambda _6$  in equation (\ref{musaa}) we get
\begin{eqnarray}
\mu_x+(\frac{2-\dot{x}t^2-4tx+t^2x^2} {t(2-tx)}\mu)_{{\dot{x}}} = 0.
\label{mu1eq11}
\end{eqnarray}
Eq. (\ref{mu1eq11}) admits a particular solution of the form
\begin{eqnarray}
\mu_2=-\frac{3 t(2-tx)}
{(2-2 tx+t^2\dot{x}+t^2x^2)^2}.
\label{mu1eq33}
\end{eqnarray}
We find that $\mu_2$ also satisfies the first equation given in (\ref{musaa}) and forms a compatible solution to the system of equations (\ref{musaa}).

Substituting the expressions $\lambda_6$ and $\mu_2$ in (\ref{imueq}) and integrating the resultant set of equations, 
\begin{eqnarray}
I_t & = -\frac{3(2tx^3-t^2x^4+2(1+tx
-t^2{x}^2)\dot x-t^2 \dot x^2)}
{(2-2tx+t^2\dot{x}+t^2x^2)^2},
\nonumber\\
I_x & = -\frac{3(2+t^2x^2-4tx
-t^2\dot{x})}
{(2-2tx+t^2\dot{x}+t^2x^2)^2},
\nonumber\\
I_{\dot{x}} & = -\frac{3t(2-tx)}
{(2-2tx+t^2\dot{x}+t^2x^2)^2},
\label{me1eq3}
\end{eqnarray}
we find $3\tilde{I_2}=\hat{I_2}$, where $\hat{I_2}$ is given in (\ref{i12}). With the help of $\hat{I_1}$ and $\tilde{I_2}$, we can derive the general solution for MEE.

\section{Hidden Symmetries}
\label{hidd}
An important application of Lie point symmetry is that it can be used to reduce the order of the underlying ordinary differential equation. It was observed that the order of the reduced ODE may admit more or lesser number of symmetries than that of the higher order equation. Such symmetries were termed as ``hidden symmetries". This type of symmetry was first observed by Olver and later extensively investigated by Abraham-Shrauner and her collaborators \cite{nonlocal1,nonlocal2,nonlocal3,nonlocal4,nonlocal5,hiden_not}.

The motivation for finding hidden symmetries of differential equations is the possibility of transforming a given ODE which has insufficient number of Lie point symmetries to be solved to another ODE that has enough Lie point symmetries such that it can be solved by integration.  These hidden symmetries cannot be found through the Lie classical method for point symmetries of differential equations.  

A detailed study on the hidden symmetries of differential equations show that there can be two types of hidden symmetries. For example, if a $n^{th}$ order ODE is reduced in order by a symmetry group then two possibilities may occur. The reduced lower order ODE may not retain other symmetry group of the $n^{th}$ order ODE. Here the symmetries of the $n^{th}$ order equation is lost in the reduced equation. This lost symmetry is called a Type I hidden symmetry of the lower order ODE. Conversely, the lower order ODE may possess a symmetry group that is not shared by the $n^{th}$ order ODE. In this case, the lower order ODE has gained one symmetry. This is called as Type II hidden symmetry of the $n^{th}$ order ODE \cite{nonlocal1,nonlocal2,nonlocal3,nonlocal4,ci1,hiden_not}.

Since we are focusing our attention on second-order ODEs, we again consider the MEE equation as an example and point out the hidden symmetries associated with this equation.


\subsection{Example: modified Emden equation}
Let us consider the MEE equation (\ref{main}). By introducing the following Riccati transformation
\begin{eqnarray}
 x=\frac{\dot{y}}{y},~~t=z,\label{hid_tra}
\end{eqnarray}
the MEE can be transformed to a linear third-order ODE, that is $\frac{d^3y}{dz^3}=0$. The transformation (\ref{hid_tra}) is nothing but the invariants associated with the Lie point symmetry $y\frac{\partial}{\partial y}$ of the linear third-order ODE. In other words, the third-order equation $\frac{d^3y}{dz^3}=0$ has been order reduced to MEE by one of its point symmetry generator $y\frac{\partial}{\partial y}$. The other Lie point symmetries of the third-order linear ODE, $\frac{d^3y}{dz^3}=0$, are \cite{olv,nonlocal,hiden_not}
\begin{eqnarray}
&& \chi_1=\frac{\partial}{\partial z},\,\chi_2=\frac{\partial}{\partial y},\,\chi_3=z^2\frac{\partial}{\partial y},\,\chi_4=z\frac{\partial}{\partial z},\nonumber\\
&&\chi_5=z\frac{\partial}{\partial y},\,\chi_6=y\frac{\partial }{\partial y},\,\chi_7=\frac{z^2}{2}\frac{\partial}{\partial z}+yz\frac{\partial}{\partial y}.\label{third_free}
\end{eqnarray}

 Substituting the transformation (\ref{hid_tra}) in the remaining vector fields given in (\ref{third_free}), they can be transformed into the following forms, namely
\begin{eqnarray} 
\hspace{-1.2cm}&&\hat{V}_1=\frac{\partial} {\partial t}=V_1,~~\hat{V}_2=-xe^{-\int x dt}\frac{\partial} {\partial x},~~\hat{V}_3=\left(\frac{t}{x}-\frac{t^2} {2}\right)xe^{-\int xdt}
\frac{\partial}{\partial x},\nonumber\\
\hspace{-1.2cm}&&\hat{V}_4=t\frac{\partial} {\partial t}-x\frac{\partial} {\partial x}=V_2-V_5,~~\hat{V}_5=\left(\frac{1}{x}-t\right)xe^{-\int xdt} \frac{\partial}{\partial x},\nonumber\\
\hspace{-1.2cm}&&
\hat{V}_7=\frac{t^2} {2}\frac{\partial}{\partial t}+(1-tx)\frac{\partial}{\partial x}=V_7-V_6,\label{hid_sym}
\end{eqnarray}
where $\hat{V}_i,~i=1,2,3,4,5,7,$ are the symmetry generators of the MEE (see Eqs.(\ref{vf8}) and (\ref{sym12})). While three of the vector fields ($\hat{V_1}$, $\hat{V_4}$ and $\hat{V_7}$) retain their point-symmetry nature, the remaining three vector fields ($\hat{V}_3, \hat{V}_3$ and $\hat{V}_5$) turns out to be nonlocal vector fields. All these vector fields satisfy the invariance condition and turn out to be the vector fields of the MEE. The local vector fields $\hat{V_1}(=V_1)$, $\hat{V_4}(=V_2-V_5)$ and $\hat{V_7}(=V_7-V_6)$ match with the earlier ones (see Eq. (\ref{vf8})) whereas the nonlocal ($\hat{V}_3, \hat{V}_3$ and $\hat{V}_5$) vector fields emerge as new ones.

Now we pick up Type-I and Type-II hidden symmetries from them. As we pointed out in the beginning of this section, Type-II hidden symmetries of third-order ODEs are nothing but the symmetries gained by the second-order ODEs. The MEE admits eight Lie point symmetries (see Sec.\ref{2}). In the above, we obtained only three Lie point symmetries of Eq.(\ref{main}). The remaining five Lie point symmetries are Type II hidden symmetries of the third-order ODE. These five symmetries can be gained from either non-local symmetries or contact symmetries of the third-order ODE. 

Type-I hidden symmetries of MEE are the symmetries which may not retain the symmetry group of the third-order ODE. In the present case, they turned out to be $\chi_3$ and $\chi_5$ since these two vector fields cannot be found in (\ref{hid_sym}).


\section{Nonlocal symmetries}
\label{nongan}
The study of hidden symmetries of ODEs brought out a new result. Besides point and contact symmetries, the ODEs do admit nonlocal symmetries (the symmetry is nonlocal if the coefficient functions $\xi$ and $\eta$ depend upon an integral). The associated vector field is of the form $V=\xi(t,x,\int u(t,x) dt)\frac{\partial}{\partial t}+\eta(t,x,\int u(t,x) dt)\frac{\partial}{\partial x}$. Subsequently attempts have been made to determine nonlocal symmetries of ODEs. However, due to the presence of nonlocal terms, these nonlocal symmetries cannot be determined completely in an algorithmic way as in the case of Lie point symmetries. The determination of nonlocal symmetries for second-order ODEs was initiated by Govinder and Leach \cite{nonlocal4}.  Their approach was confined to the determination of these nonlocal symmetries that reduce to point symmetries under reduction of order by $\frac{\partial}{\partial t}$. Later several authors have studied nonlocal symmetries of nonlinear ODEs \cite{nonlocal1,nonlocal2,nonlocal3,nonlocal4,nonlocal5,nonlocal7}.  Nucci and Leach have introduced a way to find the nonlocal symmetries \cite{nonlocal5}.  In the following, we present a couple of methods which determine nonlocal symmetries associated with the given equation. We again consider MEE as an example in both the methods and derive the nonlocal symmetries of it. We also discuss the connection between $\lambda$-symmetries and nonlocal symmetries.

\subsection{Method of Bluman et al. \cite{bl_p}}
In this method one essentially introduces an auxiliary ``covering'' system with auxiliary dependent variables.  A Lie symmetry of the auxiliary system, acting on the space of independent and dependent variables of the given ODE as well as the auxiliary variables, yields a nonlocal symmetry of the given ODE if it does not project to a point symmetry acting in its space of the independent and dependent variables. This method was first initiated by Bluman \cite{bl_p} and later extensively investigated by Gandarias and her collaborators \cite{si11,si12, sir2}.

Let the given second-order nonlinear ODE be of the form (\ref{main1}). To derive nonlocal symmetries of this equation, the authors introduced an auxiliary nonlocal variable $y$ with the following auxiliary system \cite{si11,si12, sir2}, 
\begin{equation}
\ddot{x}-\phi(t,x,\dot{x})=0,\;\;\dot{y}=f(t,x,y).
\label{eq2}
\end{equation}
 Any Lie group of point transformation $V=\xi(t,x,y)\frac{\partial}{\partial t}+\eta(t,x,y) \frac{\partial}{\partial x}+\psi(t,x,y)\frac{\partial}{\partial y}$, admitted by (\ref{eq2}) yields a nonlocal symmetry of the given ODE (\ref{main1}) if the infinitesimals $\xi$ or $\eta$ depend
explicitly on the new variable $y$, that is if the following condition is satisfied $\xi_y^2+\eta_y^2 \neq 0$.
As the local symmetries of (\ref{eq2}) are nonlocal symmetries of (\ref{main1}) this method provides an algorithm to derive a class of nonlocal symmetries for the given equation.  These nonlocal symmetries can be profitably utilized to derive the general solution for the given
equation. Using this procedure, Gandarias and her collaborators have constructed nonlocal symmetries for a class of equations \cite{si11,si12,sir2}.

In the following, using the ideas given above, we derive nonlocal symmetries for the MEE. 
\subsubsection{Example: modified Emden equation}
We introduce a nonlocal variable $y$ and rewrite Eq. (\ref{main}) in the form \cite{sir2}
\begin{eqnarray}
\ddot{x}+3x \dot{x}+x^3=0,~~\dot{y}=f(t,x,y),\label {sents}
\end{eqnarray}
where $f(t,x,y)$ is an arbitrary function to be determined.  Any Lie group of point transformation $V=\xi(t,x,y)\frac{\partial}{\partial t}+\eta(t,x,y) \frac{\partial}{\partial x}+\psi(t,x,y)\frac{\partial}{\partial y}$ admitted by (\ref{sents}) yields a nonlocal symmetry of the ODE (\ref{main}), if the infinitesimals $\xi$ and $\eta$  satisfy the equation  $\label{cod}\xi_y^2+\eta_y^2 \neq 0$.

The invariance of the system (\ref{sents}) under a one parameter Lie group of point transformations leads to the following set of determining equations $\xi$, $\eta$ and $f$, namely
\small
\begin{eqnarray}
\hspace{-2cm}\xi_{xx}= 0,\;\;\psi_x-f\xi_x = 0,\;\;\eta_{xx}-f_x \xi_y-2 \xi_{tx}-2 f \xi_{xy}+6  x \xi_{x} &=& 0,\nonumber\\
\hspace{-2cm}\psi_t+f \psi_y-f\xi_t-f^2\xi_y -f_t \xi-f_x \eta&=& 0,\nonumber \\ 
\hspace{-2cm}2 x^{3} \xi_t+2 f x^{3} \xi_y- \eta_x x^{3}+3 \eta x^{2}+3 \eta_t x+3 f \eta_y x +\eta_{tt}+f^{2} \eta_{yy} +2 f \eta_{yt}+f_t \eta_y&=& 0,\nonumber\\ 
\hspace{-2cm}3 x \xi_t-\xi_{tt} -f^{2} \xi_{yy}-2 f \xi_{yt} +3 f x \xi_{y}\label{ed} -f_t \xi_y+3 x^{3} \xi_{x}+f_x \eta_y+2 \eta_{tx}+2 f \eta_{xy}+3 \eta &=& 0.
\label{sym1}
\end{eqnarray}
\normalsize

Solving the overdetermined system (\ref{sym1}) we obtain the following infinitesimal symmetry generator for the Eq.(\ref{sents}):
\begin{equation}
V=c(t)e^{y}(x \frac{\partial}{\partial x}+ \frac{\partial}{\partial y}),\label{ght}
\end{equation}
with 
\begin{equation}\label{ff}
f(t,x)= -x-\frac{c_t}{c},
\end{equation}
where $c(t)$ is an arbitrary function of $t$. We note here that (\ref{ght}) is not the only solution set for the determining equation (\ref{sym1}).

Solving the characteristic equation, we find two functionally independent invariants which are of the form
\begin{equation}
\label{inv1}\begin{array}{ll}z=t, &
\zeta=\displaystyle \frac{\dot{x}}{x}+x.
\end{array}\end{equation} 
In terms of these two variables, $z$ and $\zeta$, Eq.(\ref{main}) reads as $\zeta_z+\zeta^2=0$. The general solution of this first-order ODE can be given readily as $\zeta=\displaystyle
\frac{1}{t+k_1}$ with $k_1$ as an integration constant.  Plugging this expression in the second equation in (\ref{inv1}) and rewriting it, we find
\begin{equation}
\frac{\dot{x}}{x}+x-\frac{1}{t+k_1}=0.
\end{equation}
  This first-order ODE can be integrated straightforwardly  to yield
 \begin{equation}\label{sol2}x= {{2\,\left(t+{k_1}\right)}\over{t^{2}+2 { k_1}\,t-2\,
 { k_2}}}, \end{equation}
where $k_2$ is the second integration constant. Replacing $k_1=I_1$ and $-2k_2=I_2$ in (\ref{sol2}), we end up at the expression given in Eq.(\ref{lie_soln}).

\subsection{Connection between nonlocal symmetries and $\lambda$-symmetries}
\label{mjk}
The exponential nonlocal symmetries admitted by (\ref{sents}) can also be derived from $\lambda$-symmetries. To illustrate this we recall the following theorem from Ref.\cite{tel2}. 
\begin{thmn}\label{teocasoparticular}
Let us suppose that for a given second-order equation (\ref{main1}) there exists some function $f=f(t,x,\dot{x})$ such that the system (\ref{eq2}) admits a Lie point symmetry  $V=\xi(t,x,y)\frac{\partial}{\partial t}+\eta(t,x,y) \frac{\partial}{\partial x}+\psi(t,x,y)\frac{\partial}{\partial y}$ satisfying  $\xi_y^2+\eta_y^2 \neq 0$. We assume that $z=z(t,x)$, $\zeta=\zeta(t,x,\dot{x})$ are two functionally independent functions that verify $V(z)=0, \left. V^{(1)}(\zeta)\right|_\Delta=0$ and are such that equation (\ref{main}) can be written in terms of
$\{z,\zeta,\zeta_z\}$ as a~first-order ODE.
Then
\begin{enumerate}\itemsep=0pt
\item[$1.$] The vector field $V$ has to be of the form
\begin{eqnarray}\label{final0}
V=e^{Cy}\left(\xi(t,x)\partial_t+\eta(t,x)\partial_x+\psi(t,x,\dot{x})\partial_y\right)+C_1\partial_y,
\end{eqnarray}
where $C$ and $C_1$ are constants.
\item[$2.$] The pair
\begin{eqnarray}\label{pair2}
v=
\xi(t,x)\partial_t+\eta(t,x)\partial_x,\qquad
\lambda=C f.
\end{eqnarray}
defines a $\lambda$-symmetry of the equation (\ref{main}) and the set $\{z,\zeta,\zeta_z\}$ is a complete system of invariants of $v^{[\lambda,(1)]}$.
\end{enumerate}
\end{thmn}
With the choice $C=1, C_1=0$ and $f=\lambda$, the vector field (\ref{final0}) turns out to be
\begin{eqnarray}\label{expteo}
V=e^{y}\left(\xi(t,x)\partial_t+\eta(t,x)\partial_x+\psi(t,x,\dot{x})\partial_y\right),
\end{eqnarray}
where $\xi$ and $\eta$ are the infinitesimal coefficients of $v$ and $\psi = \psi(t, x, \dot{x})$ satisfy the condition $
V^{(2)}(\dot{y} - \lambda)|_\Delta = 0$. This equation provides a linear first-order partial differential equation to
determine $\psi$, that is
\begin{eqnarray}\label{edppsi}
\psi_t+\dot{x}\psi_x +\psi_{\dot{x}}\phi+\psi\lambda=D_t(\xi)\lambda+\xi \lambda^2+v^{[\lambda,(1)]}(\lambda).
\end{eqnarray}

Let $v=\xi \partial_t+\eta \partial_x$ be a $\lambda$-symmetry of (\ref{main1}) for some $\lambda=\lambda(t,x,\dot{x})$ and $\psi=\psi(t,x,\dot{x})$ be a particular solution of the equation (\ref{edppsi}).  Then (\ref{expteo})
is a~nonlocal symmetry of (\ref{main1})
associated to system (\ref{eq2}) for $f=\lambda(t,x,\dot{x})$ \cite{tel2} .

\subsubsection{Example: modified Emden equation}
Using the above, we can demonstrate that the nonlocal symmetries found by Gandarias et.al for the MEE can be extracted from the $\lambda$-symmetries themselves. To show this let us consider the $\lambda$-symmetry $\frac{\partial} {\partial x}$ with $\lambda_3=\frac{\dot{x}} {x}-x$ (from Table 1). Substituting this expression in Eq.(\ref{edppsi}) and solving the resultant partial differential equation we can obtain an explicit expression for $\psi(t,x,\dot{x})$. Let us choose the simplest case $\psi(t,x,\dot{x})=0$. In this case the left hand side of Eq.(\ref{edppsi}) disappears and the right hand side of it also vanishes automatically since it is nothing but the $\lambda$-symmetry determining equation in which $\lambda_3$ is a solution. Substituting $\lambda_3=f$ in the second expression given in (\ref{eq2}) and integrating it, we find
\begin{equation}
y=\log x-\int x dt.
\end{equation}
Now substituting the expressions $\xi=0$, $\eta=1$, $\psi=0$ and the above expression of $y$ in (\ref{expteo}), we obtain a nonlocal symmetry
\begin{eqnarray}
\Omega_4=xe^{-\int xdt}\frac{\partial}{\partial x}.\label{som_new}
\end{eqnarray}
One can unambiguously verify that the vector field (\ref{som_new}) also satisfies the determining equation and turns out to be a nonlocal symmetry of the MEE. We mention here that the nonlocal symmetry (\ref{som_new}) had already been observed in the order reduction procedure (see Eq. (\ref{hid_sym})). 

The other choices of $\lambda$ and/or $\psi(t,x,\dot{x})$ will generate new nonlocal symmetries for the MEE.  For example, the choice $\psi=c(t),\xi=0,\eta=c(t)x$ and $\lambda=-x-\frac{c_t}{c}$, provides another nonlocal symmetry (\ref{ght}) through the above said procedure. 
In this way, one can also construct nonlocal symmetries from the $\lambda$-symmetries. 

\subsection{Method of Gladwin Pradeep et.al}
\label{anna_work}  
In a recent paper Gladwin Pradeep et.al proposed yet another procedure to determine the nonlocal symmetries for the given equation \cite{nonlocal}. In the following, we briefly recall the essential ideas behind this method with reference to MEE.

The MEE (\ref{main}) can be transformed to the second-order linear ODE $\frac{d^2u} {dt^2}=0$
through the nonlocal transformation $u=xe^{\int xdt}$.
To explore the nonlocal symmetries associated with (\ref{main}), the authors used the identity $\frac{\dot{u}} {u}=\frac{\dot{x}} {x}+x$ (which can be directly deduced from the nonlocal transformation $u=xe^{\int xdt}$) \cite{nonlocal}. This nonlocal connection between the free particle equation and MEE allows one to deduce the nonlocal symmetries of Eq. (\ref{main}) in the following manner.
 
Let $\xi$ and $\eta$ be the infinitesimal point transformations, that is $u'=u+\epsilon \eta(t,u)$, $T=t+\epsilon \xi(t,u)$, associated with the linear ODE $\frac{d^2u} {dt^2}=0$.  The symmetry vector field associated with this infinitesimal transformations reads as $V=\xi \frac{\partial}{\partial t}+\eta \frac{\partial}{\partial u}$
and its first extension is given by $Pr^{(1)}V=\xi \frac{\partial}{\partial t}
+\eta \frac{\partial}{\partial u}+(\dot{\eta}
-\dot{u}\dot{\xi})\frac{\partial}{\partial \dot{u}}$. Then we denote the symmetry vector field and its first prolongation of the MEE (\ref{main}) as $\Omega=\delta \frac{\partial}{\partial t}+\mu \frac{\partial}{\partial u}$ and $Pr^{(1)}\Omega=\delta \frac{\partial}{\partial t}
+\mu \frac{\partial}{\partial x}+(\dot{\mu}
-\dot{x}\dot{\delta})\frac{\partial}{\partial \dot{x}}$, where $\delta$ and $\mu$ are the infinitesimals associated with the variables $t$ and $x$, respectively.  

From the identity $\frac{\dot{u}} {u}=\frac{\dot{x}} {x}+x$, the authors defined a new function, say $X$
\begin{eqnarray}            
\frac{\dot{u}}{u}=\frac{\dot{x}}{x}+x=X.\label{x}
\label {sym05}
\end{eqnarray}
In the new variable $X$, the MEE turns out to be the Riccati equation, that is
\begin{eqnarray}
\dot{X}+X^2=0.
\label{reduced-riccati}
\end{eqnarray}

The symmetry vector field of this equation can be obtained by using the relation $X=\frac{\dot{u}}{u}$ and  rewriting $V^{1}=\xi \frac{\partial}{\partial t}
+\eta \frac{\partial}{\partial u}+(\dot{\eta}
-\dot{u}\dot{\xi})\frac{\partial}{\partial \dot{u}}$ as
\begin{eqnarray}
&& V^{1}=\xi \frac{\partial}{\partial t}
+\bigg(\frac{\dot{\eta}}{u}-\frac{\eta \dot{u}}{u^2}
-X\dot{\xi}\bigg) \frac{\partial}{\partial X}\equiv\Sigma.
\label {sym06}
\end{eqnarray}
We note that Eq. (\ref{reduced-riccati}), being a first-order ODE, admits infinite number of Lie point symmetries.  These Lie point symmetries of Eq. (\ref{reduced-riccati}) become contact symmetries of the linear second-order ODE $\frac{d^2u} {dt^2}=0$ through the relation $X=\frac{\dot{u}}{u}$.

Similarly one can rewrite $\Omega^{1}=\delta \frac{\partial}{\partial t}
+\mu \frac{\partial}{\partial x}+(\dot{\mu}
-\dot{x}\dot{\delta})\frac{\partial}{\partial \dot{x}}$, using the relation $X=\frac{\dot{x}}{x}+x$, as
\begin{eqnarray} 
&&\Omega^{1}=\delta \frac{\partial}{\partial t}
+\bigg((-\frac{1}{x^2}\dot{x}+1)\mu+(\dot{\mu}
-\dot{x}\dot{\lambda})\frac{1}{x}\bigg)\frac{\partial}{\partial X}\equiv\Xi.
\label {sym07}
\end{eqnarray}

As the symmetry vector fields $\Sigma$ and $\Xi$ correspond to the same
equation (\ref{reduced-riccati}), their infinitesimal symmetries must be equal.  Therefore, comparing equations (\ref{sym06}) and (\ref{sym07}) one obtains
\begin{eqnarray} 
&& \xi =\delta,\quad 
 \frac{\dot{\eta}}{u}-\frac{\eta \dot{u}}{u^2}
-x\dot{\xi}=(-\frac{1}{x^2}\dot{x}
+1)\mu+\dot{\mu}\frac{1}{x}.
\label {sym08}
\end{eqnarray}
Rewriting the second equation given in (\ref{sym08}) we can obtain the following first-order ODE for the unknown function $\mu$, that is
\begin{eqnarray} 
 \frac{1}{x}\dot{\mu}+(-\frac{1}{x^2}\dot{x}+1)\mu= 
 \frac{d}{dt}(\frac{\eta}{u})
-x\dot{\xi}.
\label {sym09}
\end{eqnarray}


The free particle equation $\frac{d^2u} {dt^2}=0$ admits eight Lie point symmetries which are given in Eq.(\ref{sym11}). Substituting these symmetries $(\xi_i,\eta_i)$, $i=1,2,\ldots,8$, and $u=xe^{\int xdt}$, in Eq. (\ref{sym09}),
we get the following seven first-order ODEs for $\mu$,
\numparts
\addtocounter{equation}{-1}
\label{dumm}
\addtocounter{equation}{1}
\begin{eqnarray}
&&\frac{1}{x}\dot{\mu}+(1-\frac{1}{x^2}\dot{x})\mu+(x^2+\dot{x})x^{-2}e^{-\int xdt}=0,\\
&&\frac{1}{x}\dot{\mu}+(1-\frac{1}{x^2}\dot{x})\mu-(x-tx^2-t\dot{x})x^{-2}e^{-\int xdt}=0,\\
&&\frac{1}{x}\dot{\mu}+(1-\frac{1}{x^2}\dot{x})\mu=0,\\
&&\frac{1}{x}\dot{\mu}+(1-\frac{1}{x^2}\dot{x})\mu+(x^2+\dot{x})xe^{-\int xdt}=0,\\
&&\frac{1}{x}\dot{\mu}+(1-\frac{1}{x^2}\dot{x})\mu+x=0,\\
&&\frac{1}{x}\dot{\mu}+(1-\frac{1}{x^2}\dot{x})\mu+2tx-1=0,\\
&&\frac{1}{x}\dot{\mu}+(1-\frac{1}{x^2}\dot{x})\mu+2(x^2+\dot{x})x^2e^{2\int xdt}-1=0.
\end{eqnarray}
\endnumparts
Integrating each one of the above first-order linear ODEs we can obtain the function $\mu$.  Substituting the symmetries $\delta(=\xi)$ and $\mu$ in $\Omega=\delta \frac{\partial}{\partial t}
+\mu \frac{\partial}{\partial x}$, we get following nonlocal symmetries,
\begin{eqnarray} 
&&\hspace{-0.5cm}\Omega_1=\frac{\partial}{\partial t},~~\Omega_2=\left(\frac{1}{x}-t\right)xe^{-\int xdt}
\frac{\partial}{\partial x},\nonumber \\   
&&\hspace{-0.5cm}\Omega_3=\left(\frac{t}{x}-\frac{t^2} {2}\right)xe^{-\int xdt}
\frac{\partial}{\partial x},~~\Omega_4=xe^{-\int xdt}\frac{\partial}{\partial x},
\label{omega4-eg1}\nonumber\\
&&\hspace{-0.5cm}\Omega_5=xe^{\int x dt}\frac{\partial}{\partial t}
-\left(\int x(\dot{x}+x^2)e^{\int(2x) dt}dt\right)xe^{-\int x dt}
\frac{\partial}{\partial x},\nonumber\\
&&\hspace{-0.5cm}\Omega_6=t\frac{\partial}{\partial t}
-xe^{-\int( x f_x )dt}\left(\int xe^{\int x dt}dt\right)  \frac{\partial}{\partial x},\nonumber\\
&&\hspace{-0.5cm}\Omega_7=t^2 \frac{\partial}{\partial t}
+xe^{-\int{x}dt}\left(\int (1-2tx)e^{-\int x dt}dt\right)  \frac{\partial}{\partial x},\nonumber\\
&&\hspace{-0.5cm}\Omega_8=txe^{\int x dt}\frac{\partial}{\partial t}
+xe^{-\int{x}dt}\left(\int (tx^3+(tx-1)\dot{x})
e^{\int (2x)dt}dt \right)
\frac{\partial}{\partial x}
\label {sym12}
\end{eqnarray}
of equation (\ref{main}). One may observe that some of the nonlocal vector fields $\Omega_2, \Omega_3$ and $\Omega_4$ had already been found as hidden symmetries earlier. It is a straightforward exercise to check that all these nonlocal symmetries satisfy the invariance condition $\delta \frac{\partial \phi}{\partial t}+\mu \frac{\partial \phi}{\partial x}+\mu^{(1)}\frac{\partial \phi}{\partial \dot{x}}-\mu^{(2)}=0$, where $\mu^{(1)}$ and $\mu^{(2)}$ are the first and second prolongations. We mention here that these nonlocal symmetries can also be related to $\lambda$-symmetries through the theorem given in Sec. \ref{mjk}.

To derive the general solution of the given nonlinear ODE one has to solve the Lagrange's system associated with the nonlocal symmetry. For the vector field $\Omega_4$, the underlying equation reads (Eq. (\ref{omega4-eg1})),
\begin{equation}
\frac{dt}{0}=\frac{dx}{x}=\frac{d\dot{x}}{\dot{x}-x^2}.\label{ghfss}
\end{equation}
Integrating Eq.(\ref{ghfss}), we find $u=t$ and $v=\frac{\dot{x}}{x}+x$. Following the procedure described in Sec.\ref{con_sub}, we can obtain the general solution of (\ref{main}) as in the form Eq.(\ref{lie_soln}).

\section{Telescopic vector fields}
\label{teles}
Telescopic vector fields are more general vector fields than the ones discussed so far. The Lie point symmetries, contact symmetries and $\lambda$-symmetries are all sub-cases of telescopic vector fields. A telescopic vector field can be considered as a $\lambda$-prolongation where the two first infinitesimals can depend on the first derivative of the dependent variable \cite{tel1,tel2}. In the following, we briefly discuss the method of finding telescopic vector fields for a second-order ODE. We then present the telescopic vector fields for the MEE.

Let us consider the second-order equation (\ref{main1}). The vector field 
\begin{equation}
v^{(2)}=\xi \frac{\partial}{\partial t}+\eta  \frac{\partial}{\partial x}+\zeta^{(1)} \frac{\partial}{\partial \dot{x}}+\zeta^{(2)} \frac{\partial}{\partial \ddot{x}}
\end{equation}
is telescopic if and only if \cite{tel1}
\begin{eqnarray}
 \xi=\xi(t,x,\dot{x}),~\eta=\eta(t,x,\dot{x}),~\zeta^{(1)}=\zeta^{(1)}(t,x,\dot{x})
\end{eqnarray}
with $\zeta^{(2)}$ is given by
\begin{eqnarray}
 \zeta^{(2)}=D[\zeta^{(1)}]-\phi D[\xi]+\frac{\zeta^{(1)}+\dot{x}D[\xi]-D[\eta]}{\eta-\dot{x} \xi}(\zeta^{(1)}-\phi \xi).
\end{eqnarray}

To prove that the telescopic vector fields are the more general vector fields, let us introduce two functions $g_1$ and $g_2$ in the following form, namely
\begin{eqnarray}
 \hspace{-0.3cm}g_1(t,x,\dot{x})=\frac{\zeta^{(1)}+\dot{x}\xi_t-\eta_t+\dot{x}(\dot{x}\xi_x-\eta_x)}{\eta-\dot{x}\xi},~~
 g_2(t,x,\dot{x})=\frac{\dot{x}\xi_{\dot{x}}-\eta_{\dot{x}}}{\eta-\dot{x}\xi}.\label{telg2}
\end{eqnarray}
We can rewrite the prolongations $\zeta^{(1)}$ and $\zeta^{(2)}$ using the above functions $g_1$ and $g_2$ as follows:
\begin{eqnarray}
 \zeta^{(1)}&=&D[\eta]-\dot{x}D[\xi]+(g_1+g_2 \phi)(\eta-\dot{x}\xi),\\
 \zeta^{(2)}&=&D[\zeta^{(1)}]-\phi{x}D[\xi]+(g_1+g_2 \phi)(\zeta^{(1)}-\phi\xi).
\end{eqnarray}
The relationship between telescopic vector fields and previously considered
vector fields can be given by the following expressions \cite{tel1,tel2}
\begin{eqnarray}
 \zeta^{(1)}&=&\eta^{(1)}+(g_1+g_2 \phi)(\eta-\dot{x}\xi),\label{kjdf}\\
 \zeta^{(2)}&=&\eta^{(2)}+(g_1+g_2 \phi)(\zeta^{(1)}-\phi\xi).\label{fkds}
\end{eqnarray}

In the above vector fields if we choose $g_1=g_2=0$ and $\xi_{\dot{x}}^2+\eta_{\dot{x}}^2 = 0$ we get the Lie point symmetries. The choice $g_1=g_2=0$ and $\xi_{\dot{x}}^2+\eta_{\dot{x}}^2 \neq 0$ gives the contact symmetries. To get $\lambda$-symmetries, we should choose $g_1\neq 0$ and $\xi_{\dot{x}}^2+\eta_{\dot{x}}^2 = 0$. As a consequence it can be considered as the more general vector field.
\subsection{Example: modified Emden equation}
To find the telescopic vector fields admitted by the MEE equation, we have to solve the invariance condition $v^{(2)}(\phi)=\xi \frac{\partial \phi}{\partial t}+\eta  \frac{\partial \phi}{\partial x}+\zeta^{(1)} \frac{\partial \phi}{\partial \dot{x}}+\zeta^{(2)} \frac{\partial \phi}{\partial \ddot{x}}=0$, with $\xi$ and $\eta$ are functions of $(t,x,\dot{x})$ and $\zeta^{(1)}$ and $\zeta^{(2)}$ are defined through (\ref{kjdf}) and (\ref{fkds}) respectively. Substituting Eq.(\ref{main}) in the invariance condition, we obtain
\begin{eqnarray}
-(3 \dot{x}+3x^2)\eta-3x\zeta^{(1)} -\zeta^{(2)}=0.\label{fyuuuu}
\end{eqnarray}
Solving equation (\ref{fyuuuu}), we obtain a telescopic vector field which is of the form
\begin{eqnarray}
\hspace{-0.5cm}\gamma_1&=&-\bigg(\frac{x}{\left(x^2+\dot{x}\right)^2}\bigg)\frac{\partial}{\partial x}+\bigg(\frac{x^2-\dot{x}}{\left(x^2+\dot{x}\right)^2}\bigg)\frac{\partial}{\partial \dot{x}}+\bigg(\frac{6 x \dot{x}}{\left(x^2+\dot{x}\right)^2}\bigg)\frac{\partial}{\partial \ddot{x}}.\label{tel_com11}
\end{eqnarray}
where the components are
\begin{eqnarray}
\xi_1=0,~\eta_1=-\frac{x}{\left(x^2+\dot{x}\right)^2},~\zeta_1^{(1)}=\frac{x^2-\dot{x}}{\left(x^2+\dot{x}\right)^2},~\zeta_1^{(2)}=\frac{6 x \dot{x}}{\left(x^2+\dot{x}\right)^2}.
\end{eqnarray}

A second telescopic vector field is found to be
\begin{eqnarray}
\hspace{-1cm}\gamma_2&=&-\bigg(\frac{t (2-t x)}{\left(t^2 \left(x^2+\dot{x}\right)-2 t x+2\right)^2}\bigg)\frac{\partial}{\partial x}+\bigg(\frac{t^2 \left(\dot{x}-x^2\right)+4 t x-2}{\left(t^2 \left(x^2+\dot{x}\right)-2 t x+2\right)^2}\bigg)\frac{\partial}{\partial \dot{x}}\nonumber \\
\hspace{-1cm}&&-\bigg(\frac{6 (t x-1) (t\dot{x}+x)}{\left(t^2 \left(x^2+\dot{x}\right)-2 t x+2\right)^2}\bigg)\frac{\partial}{\partial \ddot{x}}\label{tel_com21}
\end{eqnarray}
and its components are given by
\begin{eqnarray}
\hspace{-1cm}&&\xi_2=0,~\eta_2=-\frac{t (2-t x)}{\left(t^2 \left(x^2+\dot{x}\right)-2 t x+2\right)^2},~\zeta_2^{(1)}=\frac{t^2 \left(\dot{x}-x^2\right)+4 t x-2}{\left(t^2 \left(x^2+\dot{x}\right)-2 t x+2\right)^2},\nonumber \\
\hspace{-1cm}&&\zeta_2^{(2)}=-\frac{6 (t x-1) (t\dot{x}+x)}{\left(t^2 \left(x^2+\dot{x}\right)-2 t x+2\right)^2}.\label{tel_com2}
\end{eqnarray}

The invariants associated with a telescopic symmetry vector field can be derived by solving the associated characteristic equation. For the vector field $\gamma_1$, it reads
\begin{eqnarray}
\frac{dt} {0}=\frac{dx} {-\frac{x}{\left(x^2+\dot{x}\right)^2}}=\frac{d\dot{x}} {\frac{x^2-\dot{x}}{\left(x^2+\dot{x}\right)^2}}.
\end{eqnarray}
Using the procedure discussed in Sec.\ref{sol_lie}, we can integrate the above characteristic equation to obtain the integral given in Eq.(\ref{i11}). Repeating the procedure for the second telescopic vector field (\ref{tel_com21}) we end up at the second integral given in Eq.(\ref{i12}). From these two integrals we can derive the general solution of (\ref{main}).

\section{Conclusion}
\label{9th}
In this paper, we have reviewed continuous symmetries of second-order ODEs and elaborated the methods of finding them. To begin with we have considered Lie point symmetries and presented Lie's invariance analysis for a second-order ODE. To illustrate the method, we have considered the modified Emden equation (MEE) as an example. We have also discussed few applications of Lie point symmetries. We have demonstrated the connection between symmetries and conservation laws by recalling Noether's theorem. Few conserved quantities including energy have been identified for the MEE  through this theorem. We then considered the velocity dependent transformations and presented the method of finding contact symmetries for the second-order ODEs. We have also pointed out the contact symmetries of the MEE. We have also recalled hidden symmetries of the MEE. Some of them are found to be exponential nonlocal symmetries. The connection between symmetries and the integrating factors of ODEs was discussed through $\lambda$-symmetries approach and adjoint symmetries method. The method of finding $\lambda$-symmetries, adjoint symmetries, integrating factors and their associated integrals of a second-order ODE are discussed elaborately and illustrated with MEE as an example. We have also pointed out the connection between exponential nonlocal symmetries and $\lambda$-symmetries. Finally, we have considered a more generalized vector field, namely telescopic vector field and discussed the method of finding these generalized vector fields. For the MEE we have also brought out a couple of telescopic vector fields. We have also derived the general solution of MEE from each one of these symmetries. The symmetry methods presented here are all extendable to higher order ODEs. Through this review, we have emphasized the utility of symmetry analysis in solving ODEs.


\section*{Acknowledgments}
The authors wish to thank Professor M. Lakshmanan for suggesting us to write this review and his interest and overall guidance in this program on symmetries. The work of MS forms part of a research project sponsored by Department of Science and Technology, Government of India. The work of VKC forms part of a research project sponsored by INSA Young Scientist Project. RMS acknowledges the University Grants Commission (UGC-RFSMS), Government of India, for providing a Research Fellowship.

\end{document}